\documentclass[aps,twocolumn,showpacs,superscriptaddress]{revtex4}

\usepackage{graphicx}
\usepackage{bm}
\usepackage{amsmath}
\usepackage{dcolumn}%

\newcommand{\eg}{\ensuremath{e_{g}}}
\newcommand{\tg}{\ensuremath{t_{2g}}}

\newcommand{\mb}{\ensuremath{\mu_{\text{B}}}}

\newcolumntype{/}{D{/}{/}{2,2}}  
\newcolumntype{.}{D{.}{.}{0}}  

\begin{document}

\title{Density functional theory of resonant inelastic x-ray scattering in the
  quasi-one-dimensional dimer iridate Ba$_5$AlIr$_2$O$_{11}$ }

\author{D.A. Kukusta}

\affiliation{G. V. Kurdyumov Institute for Metal Physics of the
  N.A.S. of Ukraine, 36 Academician Vernadsky Boulevard, UA-03142
  Kyiv, Ukraine}

\author{L.V. Bekenov}

\affiliation{G. V. Kurdyumov Institute for Metal Physics of the
  N.A.S. of Ukraine, 36 Academician Vernadsky Boulevard, UA-03142
  Kyiv, Ukraine}

\author{V.N. Antonov}

\affiliation{G. V. Kurdyumov Institute for Metal Physics of the
  N.A.S. of Ukraine, 36 Academician Vernadsky Boulevard, UA-03142
  Kyiv, Ukraine}

\affiliation{Max-Planck-Institute for Solid State Research,
  Heisenbergstrasse 1, 70569 Stuttgart, Germany}

\date{\today}

\begin{abstract}

  We have investigated the electronic structure of Ba$_5$AlIr$_2$O$_{11}$
  within the density functional theory using the generalized gradient
  approximation while considering strong Coulomb correlations in the framework
  of the fully relativistic spin-polarized Dirac linear muffin-tin orbital
  band-structure method. We have investigated the x-ray absorption spectra,
  x-ray magnetic circular dichroism, and resonant inelastic x-ray scattering
  spectra (RIXS) at the Ir $K$, $L_3$, $M_3$, $M_5$ and O $K$ edges. The
  calculated results are in good agreement with experimental data. The RIXS
  spectrum of Ba$_5$AlIr$_2$O$_{11}$ at the Ir $L_3$ edge possesses sharp
  twelve features $\le$1.5 eV corresponding to transitions within the Ir {\tg}
  levels. The excitations located from 2 to 4 eV are due to {\tg}
  $\rightarrow$ {\eg} and O$_{2p}$ $\rightarrow$ {\tg} transitions. The high
  energy peaks situated at 5$-$11 eV appear due to charge transfer
  transitions. The theory reproduces well the shape and polarization
  dependence of the oxygen O $K$ RIXS spectrum.  We have found that the
  dependence of the RIXS spectrum at the oxygen $K$ edge on the incident
  photon energy and the momentum transfer vector {\bf Q} is much stronger than
  the corresponding dependence at the Ir $L_3$ edge.

\end{abstract}


\maketitle

\section{Introduction}

\label{sec:introd}

In recent years, 5$d$ transition-metal oxides attract wide scientific
interest. Due to the extended nature of 5$d$ wave functions and wide energy
bands these oxides were expected to be weakly correlated metals. However,
surprisingly they often possess Mott-insulating states with unusual
electronic and magnetic properties \cite{KJM+08,KOK+09}. In 5$d$ systems, 
due to the strong spin-orbit coupling (SOC) the {\tg} orbitals split into a
quartet ($J_{\rm{eff}}$ = 3/2) and a doublet ($J_{\rm{eff}}$ = 1/2)
\cite{JaKh09,CPB10,WCK+14}. When incorporated with electron correlations, SOC
can give rise to some fascinating phenomena, such as topological insulators
\cite{QZ10,Ando13,WBB14,BLD16}, Mott insulators
\cite{KJM+08,KOK+09,JaKh09,WSY10,MAV+11}, Weyl semimetals
\cite{WiKi12,GWJ12,SHJ+15}, and quantum spin liquids \cite{JaKh09,KAV14}. So
far, most research has largely focused on Ir$^{4+}$ iridium oxides with a
$t_{2g}^5$ configuration
\cite{KJM+08,KOK+09,WSY10,MAV+11,MJF+12,KKJ+13,TYM+14,TKD+15,VEG+17,HAE+18,ABK20,AKU+21}.
In these oxides, the quartet $J_{eff}$ = 3/2 is fully occupied, and the
relatively narrow $J_{eff}$ = 1/2 doublet occupied by one electron can be
split by moderate Hubbard $U_{eff}$ with opening a small band gap called
the relativistic Mott gap \cite{KJM+08,MAV+11,AUU18}.

In 5$d^4$ iridium oxides, the Ir$^{5+}$ ions realize completely filled
$J_{eff}$ = 3/2 and empty $J_{eff}$ = 1/2 submanifolds.  For a
{\tg}$^4${\eg}$^0$ electron configuration the system is expected to be
nonmagnetic in both the weakly and strongly correlated limits. In the weakly
correlated picture, when SOC dominates over Hund's coupling, {\tg} shells are
split into a fully filled $J_{\rm{eff}}$ = 3/2 shell and an empty
$J_{\rm{eff}}$ = 1/2 shell with a band gap between them, which leads to a
nonmagnetic insulating ground state.  In the strongly correlated picture, the
first two Hund's rules require each $d^4$ site to have total spin $S$ = 1 and
total orbital $L$ = 1 moments.  SOC yields a local $J$ = 0 state on every ion
with a nonmagnetic ground state \cite{ChBa11, CSZ+17}.  Although octahedrally
coordinated Re$^{3+}$, Os$^{4+}$, and Ir$^{5+}$ systems with a 5$d^4$
electronic configuration have been studied since 1960s \cite{EFL+61}, they
have been largely ignored in the literature.  Recent theoretical and
experimental studies suggest that novel magnetic states in iridates with
pentavalent ions can emerge from the competition between the exchange
interaction, noncubic crystal field, singlet-triplet splitting, and SOC
\cite{CQL+14,Kha13,TWY+15}.

Most studies of iridates have focused on two- or three-dimensional systems
\cite{book:GaDe13,TWY+15}. Quasi-one-dimensional iridates have been
investigated to a lesser extent, especially those with so called dimers,
commonly found in transition-metal materials, and those in which the average
number of electrons per transition metal is nonintegral. Dimerization in
transition metal compounds is observed in many systems, e.g., in vanadium
oxides \cite{IFT98,PBK+97}, titanates with the spinel structure \cite{KhMi05},
$\alpha$-TiCl$_3$ \cite{Oga60}, $\alpha$-MoCl$_3$ \cite{MYL+17}, Li$_2$RuO$_3$
\cite{MYS+07}, and $\alpha$-RuCl$_3$ \cite{BGY+18,BBL+18}. The hyperhoneycomb
iridate $\beta$-Li$_2$IrO$_3$ in the high-pressure phase above 4 GPa is
characterized by the formation of Ir$_2$ dimers on zigzag chains. In the dimer
phase the spin-orbital-entangled $J_{\rm{eff}}$ = $\frac{1}{2}$ states break
down, associated with the stabilization of the bonding state of neighboring
$d_{zx}$ orbitals \cite{AUU18,TKG+19,AKU+21}. A number of dimerized or cluster
4$d$ and 5$d$ compounds with intriguing properties have been synthesized
recently \cite{TII+16,NMB+16,DMO+17,NBB+18}.

Here we report the theoretical investigation of the electronic and magnetic
structures of Ba$_5$AlIr$_2$O$_{11}$. The crystal structure of this compound
consists of MO$_6$ and IrO$_6$ octahedra. The latter octahedra share a face
and develop Ir$_2$O$_9$ dimers, which spread along the $b$ axis. Therefore,
the compound would become geometrically frustrated in the presence of
antiferromagnetic (AFM) interaction. Novel properties are expected from this
structural arrangement in addition to those driven by SOC.
Ba$_5$AlIr$_2$O$_{11}$ features dimer chains of two inequivalent octahedra
occupied by tetravalent Ir$^{4+}$ (5$d^5$) and pentavalent Ir$^{5+}$ (5$d^4$)
ions, respectively. Ba$_5$AlIr$_2$O$_{11}$ is a Mott insulator that undergoes
a subtle structural phase transition near $T_S$ = 210 K and a transition to
magnetic order at $T_M$ = 4.5 K. The ferrimagnetic (FiM) state below $T_M$ is
highly anisotropic and resilient to a strong magnetic field (up to 14 T) but
is susceptible to even modest hydrostatic pressure \cite{TWY+15}. There are
anomalies in the electrical resistivity, dielectric constant, specific heat,
and lattice parameters observed at $T_S$ = 210 K. The experimental data
indicate that the degree of charge order between Ir$^{4+}$ and Ir$^{5+}$ ions
is increased below $T_S$.

In this work we consider the RIXS properties of Ba$_5$AlIr$_2$O$_{11}$. Since
the first publication by Kao {\it et al.} on NiO \cite{KCH+96}, the RIXS
method has shown remarkable progress as a spectroscopic technique to record
the momentum and energy dependence of inelastically scattered photons in
complex materials. RIXS rapidly became the forefront of experimental photon
science \cite{AVD+11,GHE+24}. It combines spectroscopy and inelastic
scattering to probe the electronic structure of materials. This method is an
element- and orbital- selective X-ray spectroscopy technique based on a
two-step, two-photon resonant process. It combines X-ray emission spectroscopy
(XES) with X-ray absorption spectroscopy (XAS) by measuring the coherent X-ray
emission at an incident X-ray photon energy within the near edge X-ray
absorption spectrum. In the first step (X-ray absorption), an electron of the
absorbing atom is resonantly excited from a core level to an empty state. The
resulting intermediate state carries a core hole with a very small
lifetime. In the second step (X-ray emission), the system radiatively decays
into a final state in which the core hole is filled by another electron
accompanied by photon-out emission. The polarization of the incoming and
outgoing light and the resonant energy are involved in the RIXS process,
making RIXS a simultaneous spectroscopy and scattering technique. RIXS has a
number of unique features in comparison with other spectroscopic
techniques. It covers a large scattering phase space and requires only small
sample volumes. It is also bulk sensitive, polarization dependent, as well as
element and orbital specific \cite{AVD+11}. A detail comparison with other
spectroscopic technics can be found in the recent review article
\cite{GHE+24}. The spectral broadening owing to a short core hole lifetime can
be reduced to produce RIXS spectra with high resolution. It permits direct
measurements of phonons, plasmons, single-magnon, and orbitons as well as
other many-body excitations in strongly correlated systems, such as cuprates,
nickelates, osmates, ruthenates, and iridates, with complex low-energy physics
and exotic phenomena in the energy and momentum space.

There has been great progress in the RIXS experiments over the past decade.
Most calculations of RIXS spectra of various materials have been performed
using the atomic multiplet approach with some adjustable parameters while the
number of first-principle calculations of RIXS spectra is extremely
limited. In this paper we report a theoretical first-principle study of the
RIXS spectra of Ba$_5$AlIr$_2$O$_{11}$. The RIXS spectra at the Ir $L_3$ edge
in Ba$_5$AlIr$_2$O$_{11}$ were measured by Wang {\it et al.}  \cite{WWK+19}
and Katukuri {\it et al.}  \cite{KLM+22}. The former authors measure the RIXS
spectrum for interband transitions inside {\tg} bands up to 1.2 eV. The latter
authors present the RIXS spectrum for a larger energy interval of 0-8 eV. Both
measurements show similar spectra at 0-1.2 eV.  Wang {\it et al.}
\cite{WWK+19} used a two-site cluster model with some adjustable parameters to
simulate the measured Ir $L_3$ RIXS spectrum. Katukuri {\it et al.}
\cite{KLM+22} to calculate the RIXS exciton energies also used a quantum
chemistry cluster model, which contains one Ir$_2$O$_9$ dimer unit, two
neighboring AlO$_4$ tetrahedra and 15 surrounding Ba$^{2+}$ ions. This cluster
was embedded in a set of point charges that reproduce the electrostatic
effects of the solid state environment. The RIXS spectra at the O $K$ edge in
Ba$_5$AlIr$_2$O$_{11}$ were measured by Katukuri {\it et al.}  \cite{KLM+22}
for two different $\pi$ and $\sigma$ polarizations up to 10 eV. They also
presented the polarization dependence of the XAS spectra at the oxygen $K$
edge.

We carry out here a detailed study of the electronic structure, XAS, XMCD
(x-ray magnetic circular dichroism), and RIXS spectra of
Ba$_5$AlIr$_2$O$_{11}$ in terms of the density functional theory (DFT). Our
study sheds light on the role of band structure effects and transition metal
5$d$ $-$ oxygen 2$p$ hybridization in the spectral properties of 5$d$
oxides. The energy band structure, the XAS, XMCD, and RIXS spectra of
Ba$_5$AlIr$_2$O$_{11}$ are investigated in the {\it ab initio} approach using
the fully relativistic spin-polarized Dirac linear muffin-tin orbital (LMTO)
band structure method. We use both the generalized gradient approximation
(GGA) and the GGA+$U$ approach to assess the sensitivity of the RIXS results
to different treatment of correlated electrons.

The article is organized as follows. The crystal structure of
Ba$_5$AlIr$_2$O$_{11}$ and computational details are presented in
Sec. II. Section III presents the electronic and magnetic structures of
Ba$_5$AlIr$_2$O$_{11}$. In Sec. IV, the theoretical investigations of the XAS,
XMCD, and RIXS spectra of Ba$_5$AlIr$_2$O$_{11}$ at the Ir $K$, $L_3$, $M_3$,
and $M_5$ edges are presented, the theoretical results are compared with the
experimental measurements. In Sec. V we present the theoretical investigations
of the XAS and RIXS spectra at the O $K$ edge. Finally, the results are
summarized in Sec. VI.

\section{Computational details}
\label{sec:details}

\subsection{X-ray magnetic circular dichroism.} 

Magneto-optical (MO) effects refer to various changes in the polarization
state of light upon interaction with materials possessing a net magnetic
moment, including rotation of the plane of linearly polarized light (Faraday,
Kerr rotation), and the complementary differential absorption of left and
right circularly polarized light (circular dichroism). In the near visible
spectral range these effects result from excitation of electrons in the
conduction band. Near x-ray absorption edges, or resonances, magneto-optical
effects can be enhanced by transitions from well-defined atomic core levels to
transition symmetry selected valence states \cite{book:AHY04}.

Within the one-particle approximation, the absorption coefficient
$\mu^{\lambda}_j (\omega)$ for incident x-ray polarization $\lambda$ and
photon energy $\hbar \omega$ can be determined as the probability of
electronic transitions from initial core states with the total angular
momentum $j$ to final unoccupied Bloch states

\begin{eqnarray}
\mu_j^{\lambda} (\omega) &=& \sum_{m_j} \sum_{n \bf k} | \langle \Psi_{n \bf k} |
\Pi _{\lambda} | \Psi_{jm_j} \rangle |^2 \delta (E _{n \bf k} - E_{jm_j} -
\hbar \omega ) \nonumber \\
&&\times \theta (E _{n \bf k} - E_{F} ) \, ,
\label{mu}
\end{eqnarray}
where $\Psi _{jm_j}$ and $E _{jm_j}$ are the wave function and the
energy of a core state with the projection of the total angular
momentum $m_j$; $\Psi_{n\bf k}$ and $E _{n \bf k}$ are the wave
function and the energy of a valence state in the $n$-th band with the
wave vector {\bf k}; $E_{F}$ is the Fermi energy.

$\Pi _{\lambda}$ is the electron-photon interaction
operator in the dipole approximation
\begin{equation}
\Pi _{\lambda} = -e \mbox{\boldmath$\alpha $} \bf {a_{\lambda}},
\label{Pi}
\end{equation}
where $\bm{\alpha}$ are the Dirac matrices and $\bf {a_{\lambda}}$ is the
$\lambda$ polarization unit vector of the photon vector potential,
with $a_{\pm} = 1/\sqrt{2} (1, \pm i, 0)$,
$a_{\parallel}=(0,0,1)$. Here, $+$ and $-$ denote, respectively, left
and right circular photon polarizations with respect to the
magnetization direction in the solid. Then, x-ray magnetic circular
and linear dichroisms are given by $\mu_{+}-\mu_{-}$ and
$\mu_{\parallel}-(\mu_{+}+\mu_{-})/2$, respectively.  More detailed
expressions of the matrix elements in the electric dipole
approximation may be found in
Refs.~\cite{GET+94,book:AHY04,AHS+04}.  The matrix elements due
to magnetic dipole and electric quadrupole corrections are presented
in Ref.~\cite{AHS+04}.

\subsection{RIXS}  

In the direct RIXS process \cite{AVD+11} the incoming photon with energy
$\hbar \omega_{\mathbf{k}}$, momentum $\hbar \mathbf{k}$, and polarization
$\bm{\epsilon}$ excites the solid from the ground state $|{\rm g}\rangle$ with
energy $E_{\rm g}$ to the intermediate state $|{\rm I}\rangle$ with energy
$E_{\rm I}$. During relaxation an outgoing photon with energy $\hbar
\omega_{\mathbf{k}'}$, momentum $\hbar \mathbf{k}'$ and polarization
$\bm{\epsilon}'$ is emitted, and the solid is in state $|f \rangle$ with
energy $E_{\rm f}$. As a result, an excitation with energy $\hbar \omega =
\hbar \omega_{\mathbf{k}} - \hbar \omega_{\mathbf{k}'}$ and momentum $\hbar
\mathbf{q}$ = $\hbar \mathbf{k} - \hbar \mathbf{k}'$ is created.  Our
implementation of the code for the calculation of the RIXS intensity uses
Dirac four-component basis functions \cite{NKA+83} in the perturbative
approach \cite{ASG97}. RIXS is the second-order process, and its intensity is
given by

\begin{eqnarray}
I(\omega, \mathbf{k}, \mathbf{k}', \bm{\epsilon}, \bm{\epsilon}')
&\propto&\sum_{\rm f}\left| \sum_{\rm I}{\langle{\rm
    f}|\hat{H}'_{\mathbf{k}'\bm{\epsilon}'}|{\rm I}\rangle \langle{\rm
    I}|\hat{H}'_{\mathbf{k}\bm{\epsilon}}|{\rm g}\rangle\over
  E_{\rm g}-E_{\rm I}} \right|^2 \nonumber \\ && \times
\delta(E_{\rm f}-E_{\rm g}-\hbar\omega),
\label{I1}
\end{eqnarray}
where the RIXS perturbation operator in the dipole approximation is given by
the lattice sum $\hat{H}'_{\mathbf{k}\bm{\epsilon}}=
\sum_\mathbf{R}\hat{\bm{\alpha}}\bm{\epsilon} \exp(-{\rm
  i}\mathbf{k}\mathbf{R})$, where $\bm{\alpha}$ are the Dirac matrices. The
sum over the intermediate states $|{\rm I}\rangle$ includes the contributions
from different spin-split core states at the given absorption edge. The matrix
elements of the RIXS process in the frame of the fully relativistic Dirac LMTO
method were presented in Ref. \cite{AKB22a}.

\subsection{Crystal structure.} 

\begin{figure}[tbp!]
\begin{center}
\includegraphics[width=0.95\columnwidth]{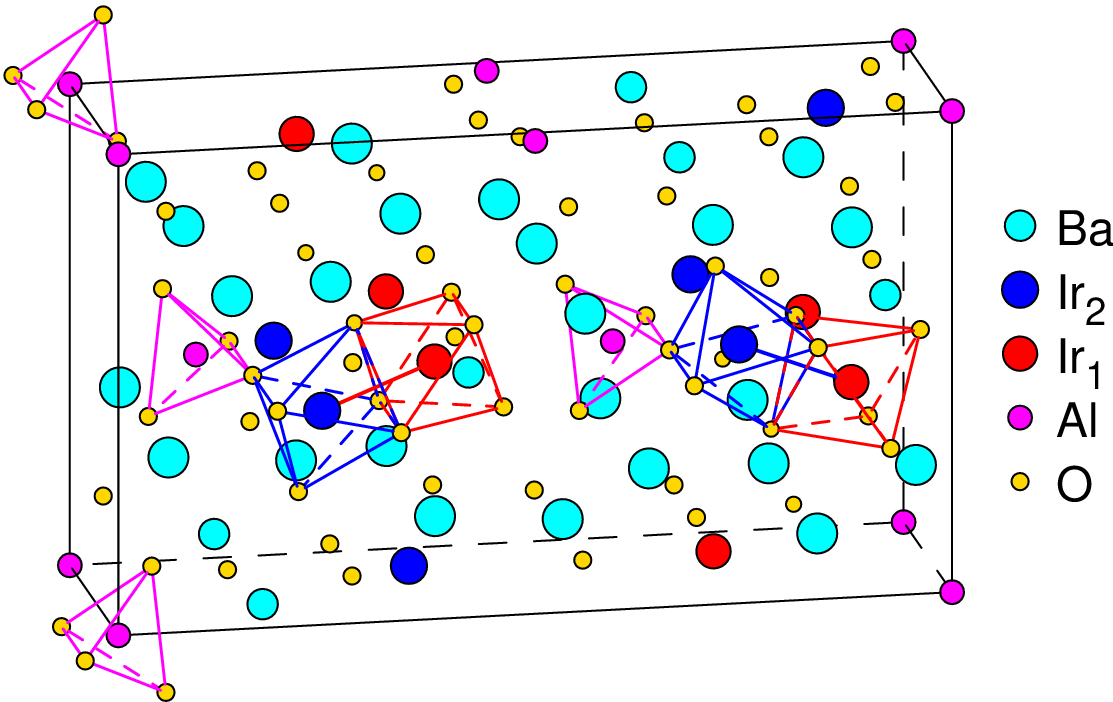}
\end{center}
\caption{\label{struc_BAIO}(Color online) The schematic
  representation of the simple orthogonal $Pnma$ (group number 62)
  Ba$_5$AlIr$_2$O$_{11}$ crystal structure \cite{MuLa89}. }
\end{figure}

Figure \ref{struc_BAIO} shows the crystal structure of Ba$_5$AlIr$_2$O$_{11}$.
It has the orthorhombic structure with space group $Pnma$ (No. 62). The
lattice parameters and Wyckoff positions are presented at Table
\ref{struc_tab_BAIO}.  The structure consists of IrO$_6$ octahedra sharing a
face along the crystallographic $b$ axis, and develop so called Ir$_2$O$_9$
dimers. Each dimer consists of two inequivalent octahedral Ir$_1$ and Ir$_2$
sites (Ir$_2$ occupies octahedra, which share their corners with AlO$_4$
tetrahedra; Ir$_1$ is in the center of the remaining octahedra) occupied by
pentavalent Ir$^{5+}$ (5$d^4$) and tetravalent Ir$^{4+}$ (5$d^5$) ions,
respectively. These dimers are corner connected through AlO$_4$ tetrahedra,
forming dimer chains along the $b$ axis, but the dimer chains are not
connected along the $a$ and $c$ axes (Fig.~\ref{struc_BAIO}).

\begin{table}[tbp!]
  \caption {The Wyckoff symbols (WS) and atomic positions ($x, y, z$) for
    Ba$_5$AlIr$_2$O$_{11}$ with the $Pnma$ structure at room temperature
    (the lattice constants $a$ = 18.8360, $b$ = 5.7887, and $c$ = 11.103 \AA)
    \cite{MuLa89}. }
\label{struc_tab_BAIO}
\begin{center}
\begin{tabular}{|c|c|c|c|c|}
\hline
         WS     & Atom     & $x$       & $y$    & $z$     \\
\hline
         $4c$   & Ba$_1$    &  0.1380   & 0.25  & 0.1920 \\
         $4c$   & Ba$_2$    &  0.9312   & 0.25  & 0.0277 \\
         $4c$   & Ba$_3$    &  0.4728   & 0.25  & 0.1048 \\
         $4c$   & Ba$_4$    &  0.1747   & 0.25  & 0.5485 \\
         $4c$   & Ba$_5$    &  0.7510   & 0.25  & 0.6230 \\
         $4c$   & Ir$_1$    &  0.5666   & 0.25  & 0.7035 \\
         $4c$   & Ir$_2$    &  0.9319   & 0.25  & 0.7062 \\
         $4c$   & Al        &  0.8110   & 0.25  & 0.2586 \\
         $4c$   & O$_1$     &  0.2540   & 0.25  & 0.365  \\
         $4c$   & O$_2$     &  0.5280   & 0.25  & 0.883  \\
         $4c$   & O$_3$     &  0.3490   & 0.25  & 0.692  \\
         $4c$   & O$_4$     &  0.2710   & 0.25  & 0.101  \\
         $4c$   & O$_5$     &  0.5950   & 0.25  & 0.538  \\
         $8d$   & O$_6$     &  0.4850   & 0.021 & 0.684  \\
         $8d$   & O$_7$     &  0.3980   & 0.482 & 0.901  \\
         $8d$   & O$_8$     &  0.3650   & 0.494 & 0.254  \\
\hline
\end{tabular}
\end{center}
\end{table}

Ba$_5$AlIr$_2$O$_{11}$ contains four formula units and hence four structural
Ir dimers in the unit cell.  The oxygen atoms surrounding the Ir sites provide
an octahedral environment. The Ir$_1-$O$_5$, Ir$_1-$O$_8$, Ir$_1-$O$_6$, and
Ir$_1-$O$_2$ interatomic distances are equal to 1.9138, 2.0196, 2.0412, and
2.1215 \AA, respectively. The corresponding distances for the Ir$_2$ ion
Ir$_2-$O$_7$, Ir$_2-$O$_3$, Ir$_2-$O$_6$, and Ir$_2-$O$_2$ are equal to
1.9047, 1.9276, 2.0600, and 2.0634 \AA\, respectively. The Al sites are
provided by an oxygen tetrahedral environment.The Al$-$O interatomic distances
are equal to 1.7314, 1.7424, and 1.7462 \AA\, for Al$-$O$_4$, Al$-$O$_1$,
Al$-$O$_8$, respectively. The Ir$_1-$Ir$_2$ intersite distance is equal to
2.72812 \AA.

Ba$_5$AlIr$_2$O$_{11}$ undergoes a subtle structural phase transition near
$T_S$ = 210 K and a magnetic order transition at $T_M$ = 4.5 K. By reducing
the temperature from 300 to 100 K the $a$, $b$, and $c$ lattice constants are
reduced approximately by 0.12, 0.22, and 0.25\%, respectively. The volume of
the unit cell is reduced from 1210.62 to 1194.84 \AA$^3$, the interatomic
Ir$_1$-Ir$_2$ distance is also reduced by approximately 0.01 \AA. All results
indicate that the phase transition at $T_S$ signals an enhanced degree of
charge order among the Ir$^{4+}$ and Ir$^{5+}$ ions \cite{MuLa89,TWY+15}.

\subsection{Calculation details}

The details of the computational method are described in our previous papers
\cite{AJY+06,AHY+07b,AYJ10,AKB22a} and here we only mention several aspects.
The band structure calculations were performed using the fully relativistic
LMTO method \cite{And75,book:AHY04}. This implementation of the LMTO method
uses four-component basis functions constructed by solving the Dirac equation
inside an atomic sphere \cite{NKA+83}. The exchange-correlation functional of
the GGA-type was used in the version of Perdew {\it et al.}
\cite{PBE96}. The Brillouin zone integration was performed using the improved
tetrahedron method \cite{BJA94}. The basis consisted of Ir, Ba, and Al $s$,
$p$, $d$, and $f$; and O $s$, $p$, and $d$ LMTO's.

To consider the electron-electron correlation effects, we used the
relativistic generalization of the rotationally invariant version of the
LSDA+$U$ method \cite{YAF03} which considers that, in the presence of SOC, the
occupation matrix of localized electrons becomes nondiagonal in spin
indexes. Hubbard $U$ was considered as an external parameter and varied from
0.65 to 3.65 eV.  We used in our calculations the value of exchange Hund's
coupling $J_H$=0.65 eV obtained from constrained LSDA calculations
\cite{DBZ+84,PEE98}. Thus, the parameter $U_{\rm{eff}}=U-J_H$, which roughly
determines the splitting between the lower and upper Hubbard bands, varied
between 0 and 3.0 eV. We adjusted the value of $U$ to achieve the best
agreement with the experiment.

In the RIXS process, an electron is promoted from a core level to an
intermediate state, leaving a core hole. As a result, the electronic
structure of this state is different from that of the ground state. To
reproduce the experimental spectrum, the self-consistent calculations
should be carried out including a core hole.  Usually, the core-hole
effect has no impact on the shape of XAS at the $L_{2,3}$ edges of
5$d$ systems and just a minor effect on the XMCD spectra at these
edges \cite{book:AHY04}. However, the core hole has a strong effect on
the RIXS spectra in transition metal compounds \cite{AKB22a,AKB22b};
therefore, we consider it.

The XAS, XMCD, and RIXS spectra were calculated taking into account the
exchange splitting of core levels. The finite lifetime of a core hole was
accounted for by folding the spectra with a Lorentzian. The widths of core
levels $\Gamma$ for Os and O were taken from Ref. \cite{CaPa01}. The finite
experimental resolution of the spectrometer was accounted for by a Gaussian of
0.6 eV (the $s$ coefficient of the Gaussian function).

Note that in our electronic structure calculations we rely on the experimentally
measured atomic positions and lattice constants because they are well
established for this material and are probably still more accurate than those
obtained from DFT.

\section{Electronic and magnetic structures}
\label{sec:bands}

The electrical resistivity $\rho$ in Ba$_5$AlIr$_2$O$_{11}$ increases by
nearly nine orders of magnitude when temperature is lowered from 750 K
(10$^2\Omega$cm) to 80 K (10$^{11}\Omega$cm). Ba$_5$AlIr$_2$O$_{11}$ is a Mott
insulator that undergoes a subtle structural phase transition near $T_S$ = 210
K and a magnetic order transition at $T_M$ = 4.5 K. $\rho$ exhibits a distinct
slope change near $T_S$ = 210 K and follows an activation law reasonably well
(better than power laws) in a temperature range of 200-750 K, which yields the
dielectric nature of Ba$_5$AlIr$_2$O$_{11}$ with an activation energy gap
$\Delta_a \sim$ 0.57 eV \cite{TWY+15}. The more rapid increase in $\rho$ below
$T_S$ indicates a charge-order transition which is accompanied by increasing
of the electronic states localization. The dielectric constant
$\varepsilon(T)$ and specific heat $C(T)$ are also consistent with a bulk
transition to long-range order at $T_S$ \cite{TWY+15}.

\begin{figure}[tbp!]
\begin{center}
\includegraphics[width=0.99\columnwidth]{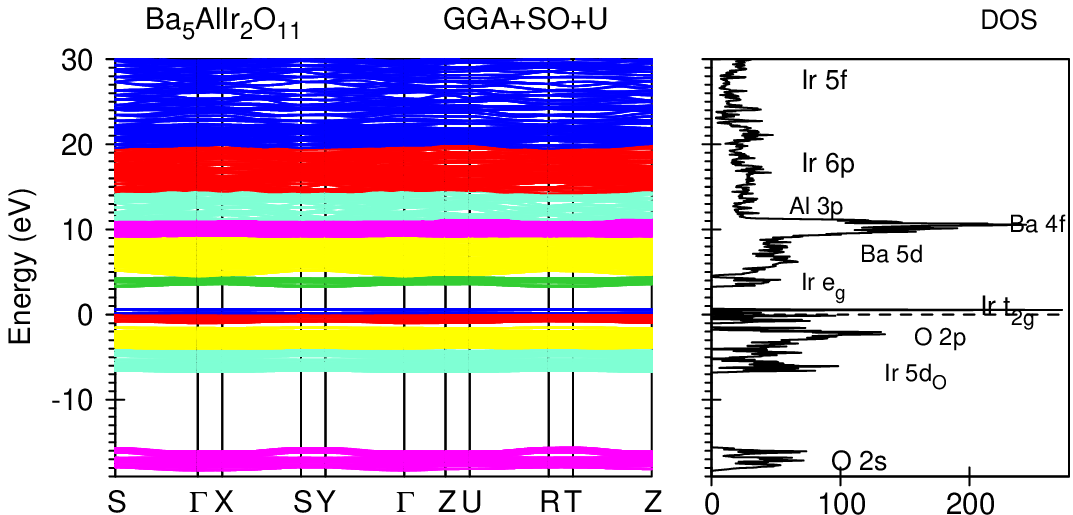}
\end{center}
\caption{\label{BND_BAIO}(Color online) The energy band structure
  of Ba$_5$AlIr$_2$O$_{11}$ calculated in the GGA+SO+$U$
  ($U_{\rm{eff}}$= 0.6 eV) approach. }
\end{figure}

We performed GGA, GGA+SO, and GGA+SO+$U$ calculations of the electronic and
magnetic structures of Ba$_5$AlIr$_2$O$_{11}$ for the experimental crystal
structure \cite{MuLa89}. Our GGA and GGA+SO band structure calculations
produce a metallic solution for Ba$_5$AlIr$_2$O$_{11}$ in contradiction to
experiment. To produce the correct dielectric ground state, one has to take
into account strong Coulomb correlations in Ba$_5$AlIr$_2$O$_{11}$. We found
that the value of Hubbard $U_{\rm{eff}}$ = 0.6 eV applied for the Ir sites
produces the best agreement between the calculated and experimentally measured
RIXS spectra at the Ir $L_3$ edge in Ba$_5$AlIr$_2$O$_{11}$. The GGA+SO+$U$
approach shifts the occupied and empty {\tg} bands downward and upward,
respectively, by $U_{\rm{eff}}$/2 producing an energy gap of 0.126 eV for
$U_{\rm{eff}}$ = 0.6 eV. The energy gap is increased with increasing Hubbard
$U$.

\begin{figure}[tbp!]
\begin{center}
\includegraphics[width=0.99\columnwidth]{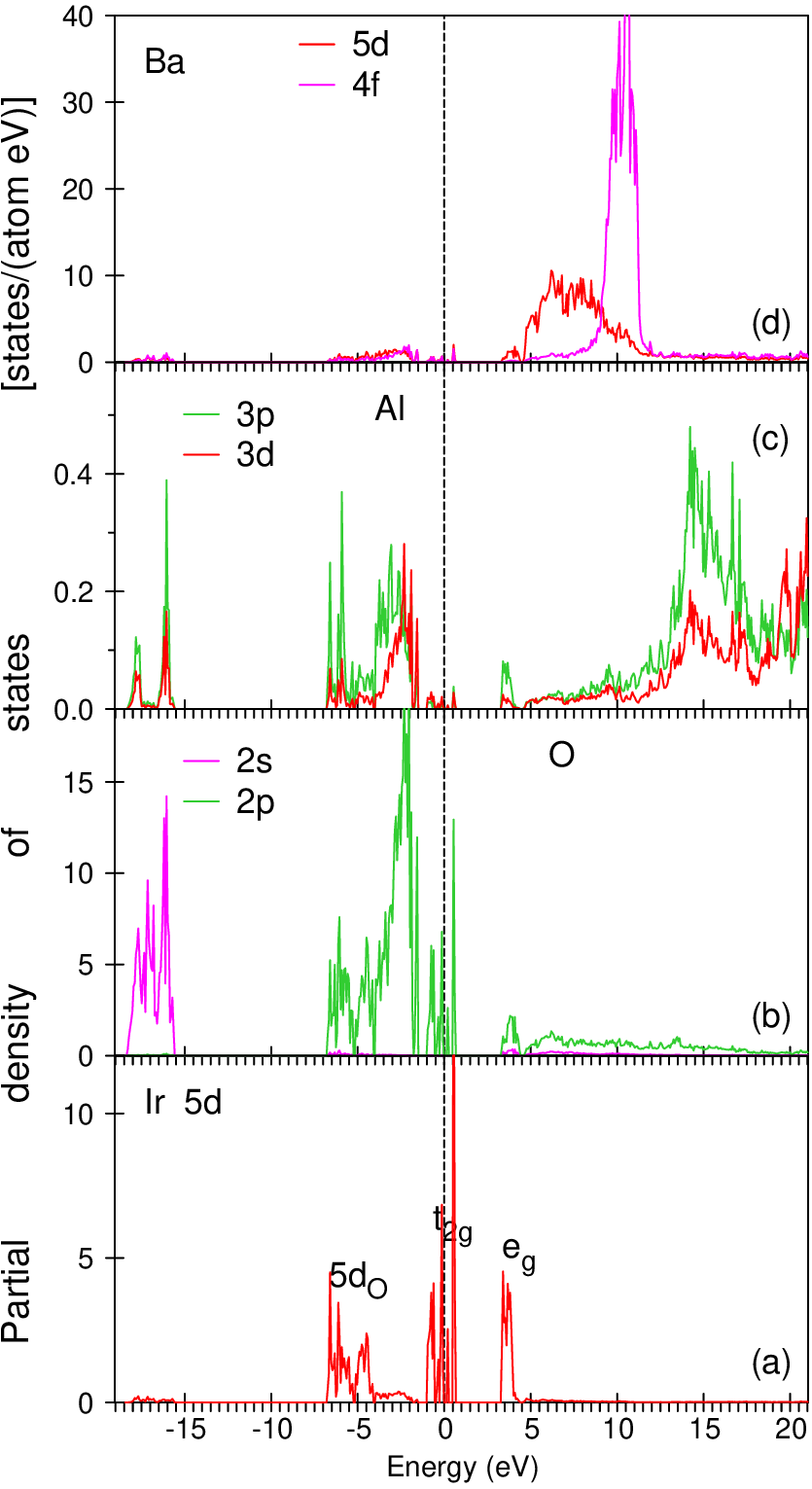}
\end{center}
\caption{\label{PDOS_BAIO}(Color online) The partial density of states (DOS)
  for Ba$_5$AlIr$_2$O$_{11}$ calculated in the GGA+SO+$U$ ($U_{\rm{eff}}$= 0.6
  eV) approach. }
\end{figure}

Figures \ref{BND_BAIO} and \ref{PDOS_BAIO} present the energy band structure
and partial density of states (DOS) in Ba$_5$AlIr$_2$O$_{11}$ calculated in
the GGA+SO+$U$ approach with $U_{\rm{eff}}$ = 0.6 eV. The occupied {\tg}
states, the low energy band (LEB), are situated in the energy interval from
$-$1 eV to $E_F$. The empty {\tg} states [the upper energy band (UEB)] consist
of three narrow single peaks divided by energy gaps and occupy the energy
range from 0.126 to 0.6 eV. The {\eg}-type states of Ir are distributed far
above the Fermi level from 3.3 to 4.3 eV. The oxygen 2$s$ states lay at the
bottom of the valence band and occupy the energy interval from $-$18.4 to
$-$15.6 eV. The oxygen 2$p$ states occupy the energy interval from $-$4.0 to
$-$1.5 eV and separate from Ir {\tg} bands by an energy gap of 0.6 eV. It is
interesting to note the existence of a strong and quite narrow O 2$p$ peak of
0.1 eV at the top of the oxygen 2$p$ band between $-$1.7 and $-$1.5 eV
separated by the rest of the 2$p$ band by a small energy gap. The oxygen 2$p$
states are strongly hybridized with the Ir {\tg} LEB and Ir {\tg} UEB just
below and above the Fermi level. The empty oxygen 2$p$ states are also
strongly hybridized with Ir {\eg} states from 3.3 to 4.3 eV
[Fig. \ref{PDOS_BAIO}(b)]. There is a significant amount of Ir 5$d$ DOS
located at the bottom of oxygen 2$p$ states from $-$6.8 to $-$4.1 eV below the
Fermi energy. The excessive charge is provided by the tails of oxygen 2$p$
states inside the Ir atomic spheres. These so called Ir 5$d_{\rm{O}}$ states
play an essential role in the RIXS spectrum at the Ir $L_3$ edge (see Sec.
IV). The Ba 5$d$ states occupy the energy region from 4.4 to 11.3 eV above the
energy Fermi. A narrow and intensive DOS peak of Ba 4$f$ states are located
just above the Ba 5$d$ states from 9.2 to 11.1 eV. The Al 3$p$ states lay
between Ba 4$f$ and Ir 6$p$ states. The latter ones occupy the energy interval
from 14.4 to 19.7 eV. Above this energy Ir 5$f$ states are situated.

Katukuri {\it et al.} \cite{KLM+22} showed that formation of molecular
orbitals is possible in Ba$_5$AlIr$_2$O$_{11}$ when the Ir$_1$-Ir$_2$
intersite distance is below 2.65 \AA. In that case, a strongly hybridized
$a_{1g}$ type of orbitals is formed and the two Ir ions in the Ir$_2$O$_9$
dimer unit (Ir$^{4+}$ and Ir$^{5+}$) preserve their local $J_{\rm{eff}}$
character close to 1/2 and 0, respectively. At room temperature the
Ir$_1$-Ir$_2$ intersite distance is equal to 2.728 \AA\, ($\sim$2.72 \AA\,
below 4 K). It is still too large for the formation of molecular orbitals in
Ba$_5$AlIr$_2$O$_{11}$. The molecular orbitals can be created by an external
or chemical pressure.

The average Ir$_1$-O bond distance $d_{\rm{Ir_1-O}}$ equals to 2.024 \AA,
while the average Ir$_2$-O bond distance is 1.989 \AA\, at room temperature.
$d_{\rm{Ir_1-O}} > d_{\rm{Ir_2-O}}$ due to different oxidation states of the
two Ir sites. $d_{\rm{Ir_1-O}}$ is longer most likely due to the relatively
large ionic radius $r$ of Ir$^{4+}$ ($r$ = 0.625 and 0.570 \AA\, for Ir$^{4+}$
and Ir$^{5+}$, respectively). Therefore, there is a tendency to charge
ordering in Ba$_5$AlIr$_2$O$_{11}$.  The charge ordering is supported by the
structural transition and anomalies in $\rho(T)$ and $C(T)$ at $T_S$
\cite{TWY+15}. However, the charge disproportionation of $\sim$0.3 electron is
not complete. Similar disproportionation was observed by Terzis {\it et al.}
\cite{TWY+15}. A purely ionic model with strong SOC, which would support
$J_{\rm{eff}}$ = $\frac{1}{2}$ in Ir$_2^{4+}$ (5$d^5$) and $J_{\rm{eff}}$ = 0
in Ir$_1^{5+}$ (5$d^4$), is not entirely applicable in
Ba$_5$AlIr$_2$O$_{11}$. Our GGA+SO+$U$ calculations produce the ionicity equal
to +4.3 and +4.7 for the Ir$_1$ and Ir$_2$ ions, respectively. We should
mention, however, that usually the DFT approach overestimates the effect of
hybridization between electronic states in crystals
\cite{book:AHY04}. Therefore, one would expect the situation just in between
the pure ionic and DFT pictures in Ba$_5$AlIr$_2$O$_{11}$.

Our calculations give a FiM solution for Ba$_5$AlIr$_2$O$_{11}$.  The
theoretically calculated spin magnetic moments are equal to $-$0.1293 and
0.4322 {\mb} for Ir$_1$ and Ir$_2$, respectively (Table \ref{mom_BAIO}).  The
orbital magnetic moment for Ir$_1$, which is supposed to be in 5$d^{4+}$
state, is equal to $M_l^{Ir_1}$ = $-$0.2511 {\mb}. For the Ir$_2^{+5}$ ion the
orbital moment is small but does not vanish $M_l^{Ir_2}$ = 0.0229 {\mb}. We
can conclude that in the DFT approach there is no pure Ir$^{4+}$ or Ir$^{5+}$
states in Ba$_5$AlIr$_2$O$_{11}$. For example, the orbital magnetic moment in
the $J_{\rm{eff}}$ = $\frac{1}{2}$ Mott isolator Sr$_2$IrO$_4$ at the
Ir$^{4+}$ site is $M_l$ = 0.4447 {\mb} \cite{AKB24} which is almost two times
larger than the corresponding orbital magnetic moment at the Ir$_1$ site in
Ba$_5$AlIr$_2$O$_{11}$.

\begin{table}[tbp!]
  \caption{\label{mom_BAIO} The theoretical spin $M_s$,
    orbital $M_l$, and total $M_{tot}$ magnetic moments ({\mb}) in
    Ba$_5$AlIr$_2$O$_{11}$ calculated in the GGA+SO+$U$ approach
    ($U_{\rm{eff}}$ = 0.6 eV). }
\begin{center}
\begin{tabular}{|c|c|c|c|}
\hline
 atom  & $M_s$ & $M_l$ &  $M_{tot}$ \\
\hline
 Ba$_1$ &  0.0071 & -0.0026 &  0.0045  \\
 Ba$_2$ &  0.0069 &  0.0061 &  0.0130  \\
 Ba$_3$ &  0.0076 & -0.0007 &  0.0068  \\
 Ba$_4$ &  0.0001 & -0.0030 & -0.0029  \\
 Ba$_5$ &  0.0085 &  0.0032 &  0.0117  \\
 Ir$_1$ & -0.1293 & -0.2511 & -0.3804  \\
 Ir$_2$ &  0.4322 &  0.0229 &  0.4551  \\
 Al     & -0.0004 & -0.0004 & -0.0008  \\
 O$_1$  & -0.0028 &  0.0003 & -0.0025  \\
 O$_2$  &  0.0243 & -0.0017 &  0.0226  \\
 O$_3$  &  0.1443 & -0.0113 &  0.1330  \\
 O$_4$  &  0.0027 &  0.0005 &  0.0032  \\
 O$_5$  & -0.0382 & -0.0668 & -0.1050  \\
 O$_6$  &  0.0193 &  0.0033 &  0.0226  \\
 O$_7$  &  0.0362 &  0.0097 &  0.0459  \\
 O$_8$  & -0.0072 &  0.0029 & -0.0043  \\
\hline
\end{tabular}
\end{center}
\end{table}

The spin and orbital magnetic moments at the Al site are relatively small
(both are equal to $-$0.0004 {\mb}). The largest oxygen spin and orbial
magnetic moments are found at the O$_3$ and O$_5$ sites (Table
\ref{mom_BAIO}). The Ba 5$d$ and 4$f$ states are almost completely empty in
Ba$_5$AlIr$_2$O$_{11}$ and the spin and orbital magnetic moments at the Ba
sites are quite small.

\section{XAS, XMCD, and RIXS spectra}
\label{sec:rixs}

\subsection{XAS and XMCD spectra at the Ir L$_{2,3}$ edges}

\begin{figure}[tbp!]
\begin{center}
\includegraphics[width=0.99\columnwidth]{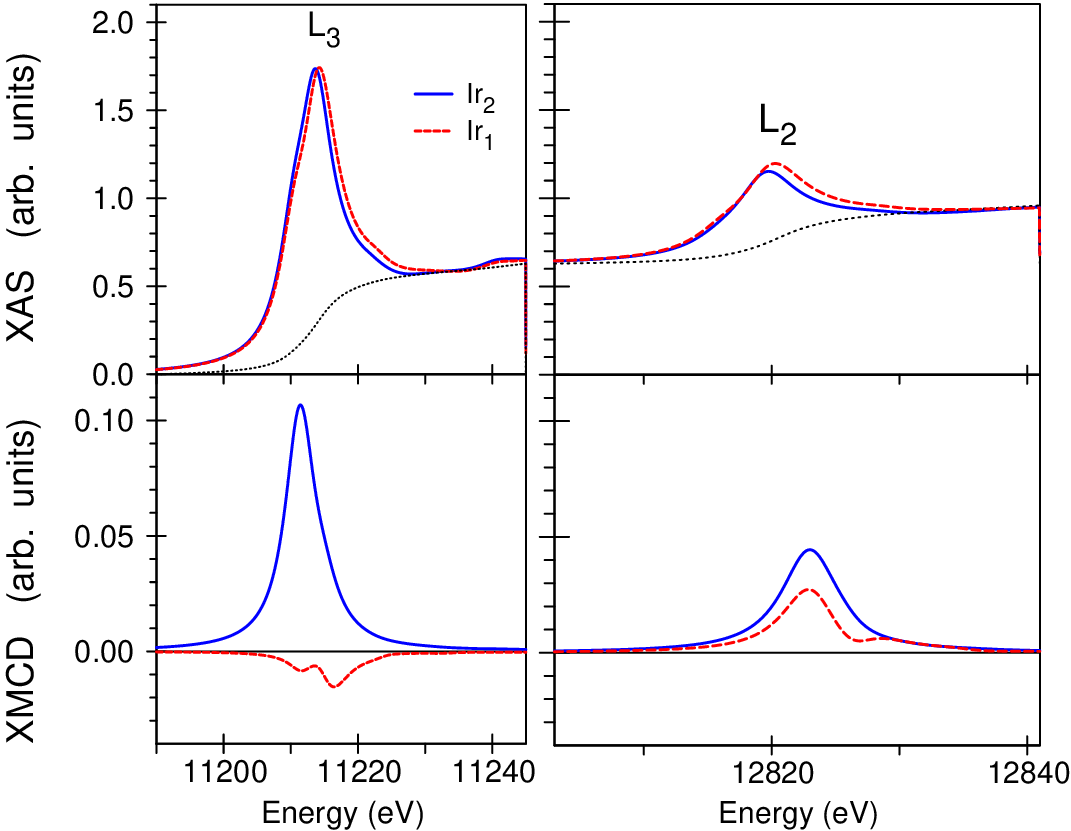}
\end{center}
\caption{\label{xmcd_Ir_BAIO}(Color online) The theoretically calculated
  x-ray absorption (the upper panels) and XMCD spectra (the lower panels) at
  the Ir $L_{2,3}$ edges in Ba$_5$AlIr$_2$O$_{11}$ in the
  GGA+SO+$U$ approach ($U_{\rm{eff}}$ = 0.6 eV). The dotted black curves in 
  the upper panels show the background scattering intensity. }
\end{figure}

Figure \ref{xmcd_Ir_BAIO} presents the XAS (the upper panel) and XMCD spectra
(the lower panel) at the Ir $L_{2,3}$ edges for Ba$_5$AlIr$_2$O$_{11}$
calculated in the GGA+SO+$U$ ($U_{\rm{eff}}$ = 0.6 eV) approach for Ir$_1$
(the full blue curves) and Ir$_2$ (the dashed red curves) sites. The isotropic
XAS spectra are dominated by empty $e_g$ states with a smaller contribution
from empty $t_{2g}$ orbitals at lower energy. The XMCD spectra, however,
mainly come from the $t_{2g}$ orbitals. This results in a shift between the
maxima of the XAS and XMCD spectra.

Due to the importance of SOC effects in iridates, it is natural to quantify
the strength of the SO interaction in these compounds. XAS provides one of the
methods for this. Van der Laan and Thole showed that the so-called branching
ratio BR = $I_{L_3}/I_{L_2}$ ($I_{L_{2,3}}$ is the integrated intensity of
isotropic XAS at the $L_{2,3}$ edges) is an important quantity assocneiated
with SOI \cite{LaTh88}. The BR is directly related to the ground-state
expectation value of the angular part of the spin-orbit coupling $<{\bf L
  \cdot S}>$ through BR = $(2 + r)/(1 - r)$, where $r$= $<{\bf L \cdot
  S}>/n_h$ and $n_h$ is the number of holes in 5$d$ states \cite{LaTh88}. As a
result, XAS provides a direct probe of SOI, which is complementary to other
techniques such as the magnetic susceptibility, electron paramagnetic
resonance, and M{\"o}ssbauer spectroscopy (which probe SOC through the value
of the Lande g-factor).  In the limit of negligible SOC effects, the
statistical branching ratio BR = 2, and the $L_3$ white line is twice the size
of the $L_2$ feature \cite{LaTh88}. The theoretically calculated BR in
Ba$_5$AlIr$_2$O$_{11}$ is 4.09 and 3.70 for the Ir$_2$ and Ir$_1$ sites,
respectively, for the GGA+SO+$U$ ($U_{\rm{eff}}$ = 0.6 eV) approach. They are
differ significantly from the statistical BR = 2 in the absence of orbital
magnetization in 5$d$ states. A strong deviation from 2 indicates a strong
coupling between the local orbital and spin moments.

There is a relatively large XMCD signal at the $L_3$ edge for the Ir$_1$
site. However, for the Ir$_2$ site the dichroism is much smaller due to
the smallness of the orbital magnetic moment at that site (see Table
\ref{mom_BAIO}). 

\subsection{RIXS spectra at the Ir L$_3$ edge}

The Ir $L_{2,3}$ RIXS spectra occur from a local excitation between filled and
empty 5$d$ states. More precisely, the incoming photon excites a 2$p_{1/2}$
core electron ($L_2$ spectrum) or a 2$p_{3/2}$ one ($L_3$ spectrum) into an
empty 5$d$ state, which is followed by the de-excitation from the occupied
5$d$ state into the core level. Because of the dipole selection rules, apart
from 6$s_{1/2}$ states (which have a small contribution to RIXS due to
relatively small 2$p$ $\rightarrow$ 6$s$ matrix elements \cite{book:AHY04})
only 5$d_{3/2}$ states occur for $L_2$ RIXS, whereas for $L_3$ RIXS
5$d_{5/2}$-states also contribute. Although the 2$p_{3/2}$ $\rightarrow$
5$d_{3/2}$ radial matrix elements are only slightly smaller than the
2$p_{3/2}$ $\rightarrow$ 5$d_{5/2}$ ones, the angular matrix elements strongly
suppress the 2$p_{3/2}$ $\rightarrow$ 5$d_{3/2}$ contribution
\cite{book:AHY04}. Therefore, the RIXS spectrum at the Ir $L_3$ edge can be
viewed as interband transitions between 5$d_{5/2}$ states.

\begin{figure}[tbp!]
\begin{center}
\includegraphics[width=0.99\columnwidth]{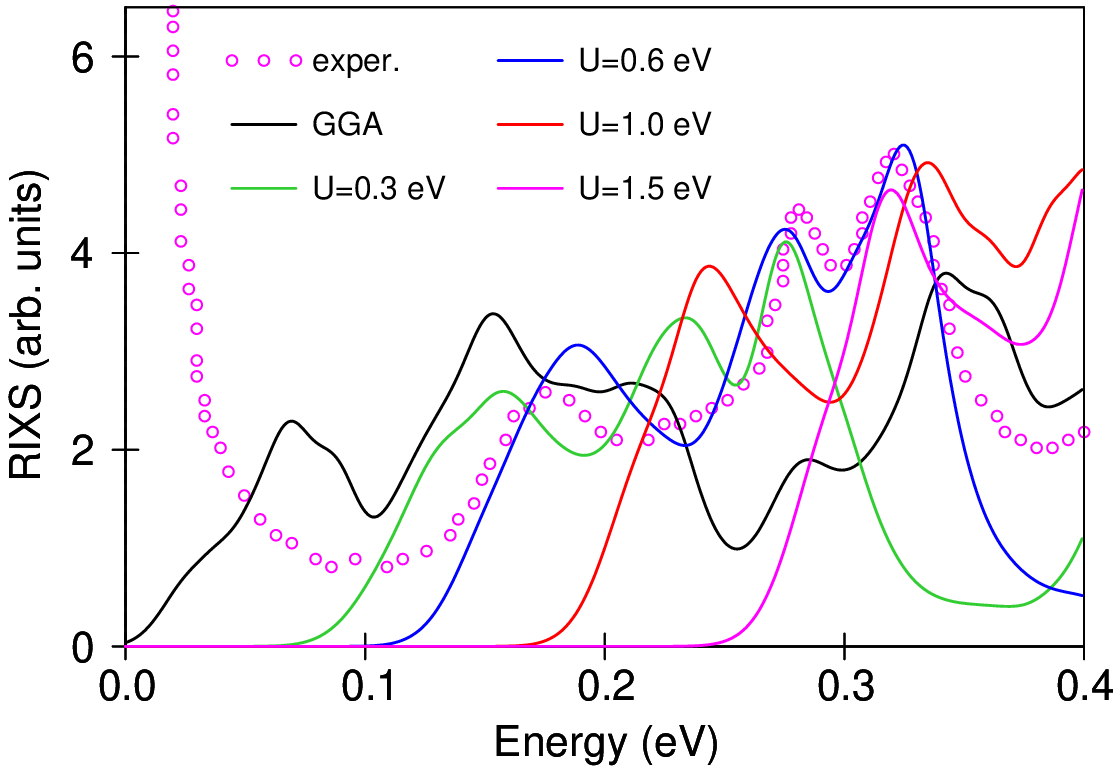}
\end{center}
\caption{\label{rixs_U_BAIO}(Color online) The experimentally measured
  RIXS spectrum at the Ir $L_3$ edge \cite{KLM+22} in
  Ba$_5$AlIr$_2$O$_{11}$ compared with the theoretical spectra
  calculated in the GGA+SO+$U$ approach for different $U_{\rm{eff}}$
  values. }
\end{figure}

\begin{figure}[tbp!]
\begin{center}
\includegraphics[width=0.9\columnwidth]{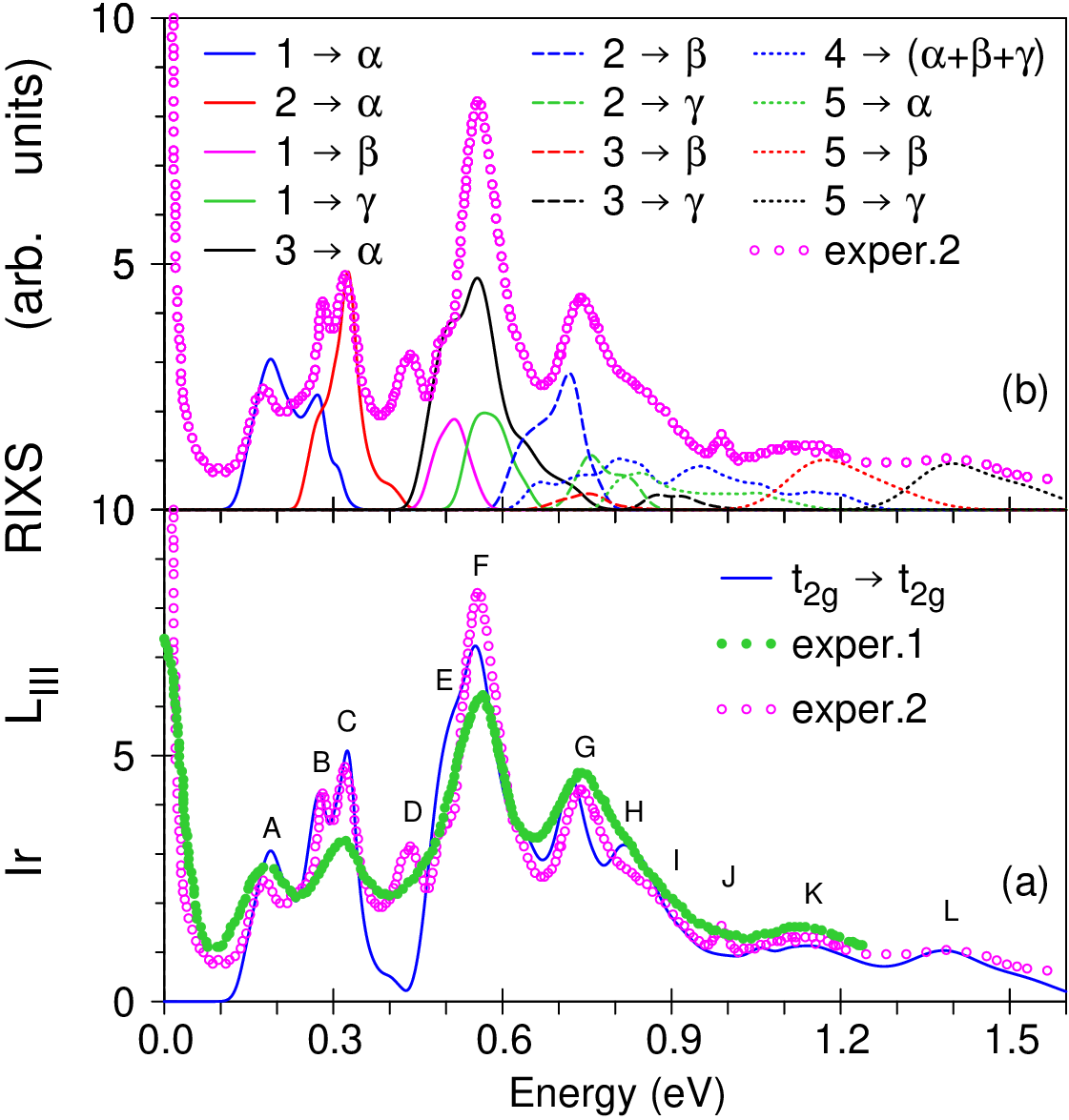}
\end{center}
\caption{\label{rixs_t2g_BAIO}(Color online) (a) The theoretical
  RIXS spectrum at the Ir $L_3$
  edge for {\tg} $\rightarrow$ {\tg} transitions in
  Ba$_5$AlIr$_2$O$_{11}$ (the full blue curve) calculated in the
  GGA+SO+$U$ approach ($U_{\rm{eff}}$=0.6 eV) compared with different
  experimental data exper.1 \cite{WWK+19} and exper.2 \cite{KLM+22}; (b)
  the experimental RIXS spectrum at the Ir $L_3$ edge for {\tg}
  $\rightarrow$ {\tg} transitions in Ba$_5$AlIr$_2$O$_{11}$ (magenta
  open circles) \cite{KLM+22} in comparison with the interband
  transitions between {\tg} states marked in
  Fig. \ref{BND_t2g_BAIO}. }
\end{figure}

Figure \ref{rixs_U_BAIO} shows the experimental RIXS spectrum measured by
Katukuri {\it et al.}  \cite{KLM+22} (open magenta circles) below 0.4 eV in
comparison with the theoretical spectra calculated in the GGA+SO and GG+SO+$U$
approximations with different values of $U_{\rm{eff}}$. The GGA+SO approach
displays unsatisfactory agreement with the experiment. The best agreement was
found for the GGA+SO+$U$ approach with $U_{\rm{eff}}$ = 0.6 eV. The
calculations with larger $U_{\rm{eff}}$ shift the RIXS spectrum toward higher
energies.

\begin{figure}[tbp!]
\begin{center}
\includegraphics[width=0.95\columnwidth]{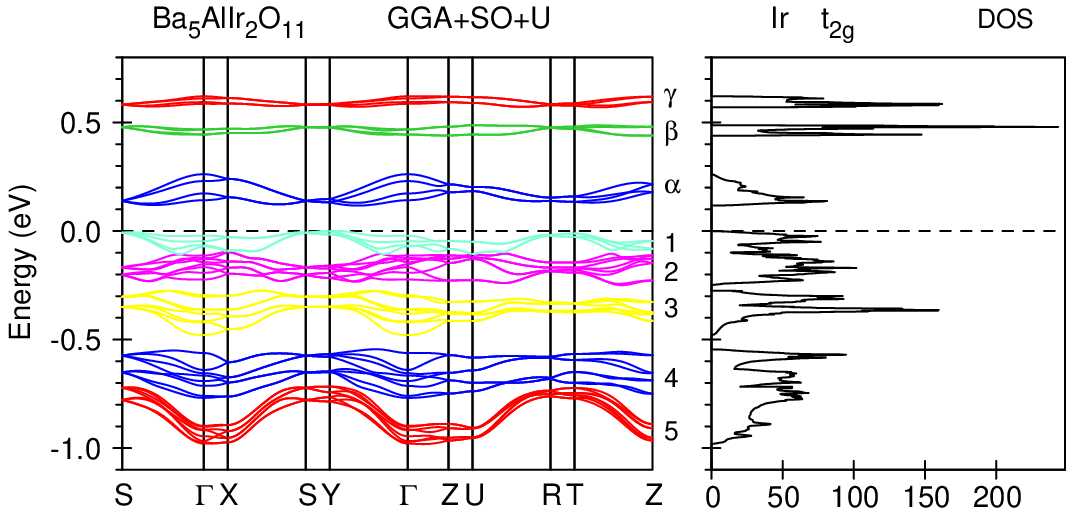}
\end{center}
\caption{\label{BND_t2g_BAIO}(Color online) The energy band structure of
  the Ir {\tg} states in Ba$_5$AlIr$_2$O$_{11}$ calculated in the GGA+SO+$U$
  ($U_{\rm{eff}}$= 0.6 eV) approach. }
\end{figure}

Figure \ref{rixs_t2g_BAIO}(a) shows the theoretical RIXS spectrum at the Ir
$L_3$ edge for {\tg} $\rightarrow$ {\tg} transitions in Ba$_5$AlIr$_2$O$_{11}$
compared with different experimental data. The measurements by Wang {\it et
  al.} \cite{WWK+19} (exper.1) and Katukuri {\it et al.} \cite{KLM+22}
(exper.2) produce quite similar RIXS spectra with just small differences in
relative peak intensities. Additionally, exper.2 possesses an extra peak $D$
and the low energy peak at $\sim$0.3 eV is split into two peaks $B$ and
$C$. The Ir $L_3$ spectrum for the {\tg} $\rightarrow$ {\tg} transitions has
quite a rich fine structure with at least twelve well separated peaks from $A$
to $L$ below 1.5 eV. This shape of the RIXS spectrum can be explained by the
specific energy band structure of Ir {\tg} states presented in
Fig. \ref{BND_t2g_BAIO}. There are five separated groups of {\tg} bands below
$E_F$ (from 1 to 5), and three groups of empty bands $\alpha$, $\beta$, and
$\gamma$, which produce three narrow DOS peaks separated by energy gaps. The
interband transitions between these five occupied and three empty groups of
bands produce 15 peaks (some of them are quite small).  As a result, we have
the rich structure of the Ir $L_3$ RIXS spectrum consisting of twelve well
distinguished peaks below 1.5 eV. The appearance of multiple peaks in the RIXS
spectrum is a direct consequence of the strong noncubic crystal field
splitting originating from the distorted octahedral environment of the Ir
ions. The corresponding interband transitions are presented in
Fig. \ref{rixs_t2g_BAIO}(b). The low energy peaks $A$, $B$, and $C$ are due to
transitions between the first (light green bands) and the second (magenta
bands) groups of occupied bands into the low energy empty bands $\alpha$ (blue
bands). The peak $D$ is absent in our calculations as well as in the
measurements of Wang {\it et al.}  \cite{WWK+19} (exper.1). The peak probably
has an exchitonic nature, as it was suggested by Katukuri {\it et al.}
\cite{KLM+22}. This needs additional theoretical consideration. The
theoretical description of exciton spectra demands a many-body approach beyond
the one-particle approximation, such as GW or Bethe-Salpeter equation
calculations. A first-principle approach for the description of exciton
spectra with the consideration of RIXS matrix elements is highly desirable.

The intensive peak $F$ at 0.55 eV with the low energy shoulder $E$ is derived
from 1 $\rightarrow$ $\beta$, 1 $\rightarrow$ $\gamma$ and 3 $\rightarrow$
$\alpha$ interband transitions. The peak $G$ is mostly due to 2 $\rightarrow$
$\beta$ interband transitions. The high energy peaks $K$ and $L$ are due to 5
$\rightarrow$ $\beta$ and 5 $\rightarrow$ $\gamma$ interband transitions,
respectively. The shoulders $H$, $I$, and $J$ occur from the combination of
many transitions such as 4 $\rightarrow$ ($\alpha$+$\beta$+$\gamma$) and 5
$\rightarrow$ $\alpha$ [see Fig. \ref{rixs_t2g_BAIO}(b)].

\begin{figure}[tbp!]
\begin{center}
\includegraphics[width=0.9\columnwidth]{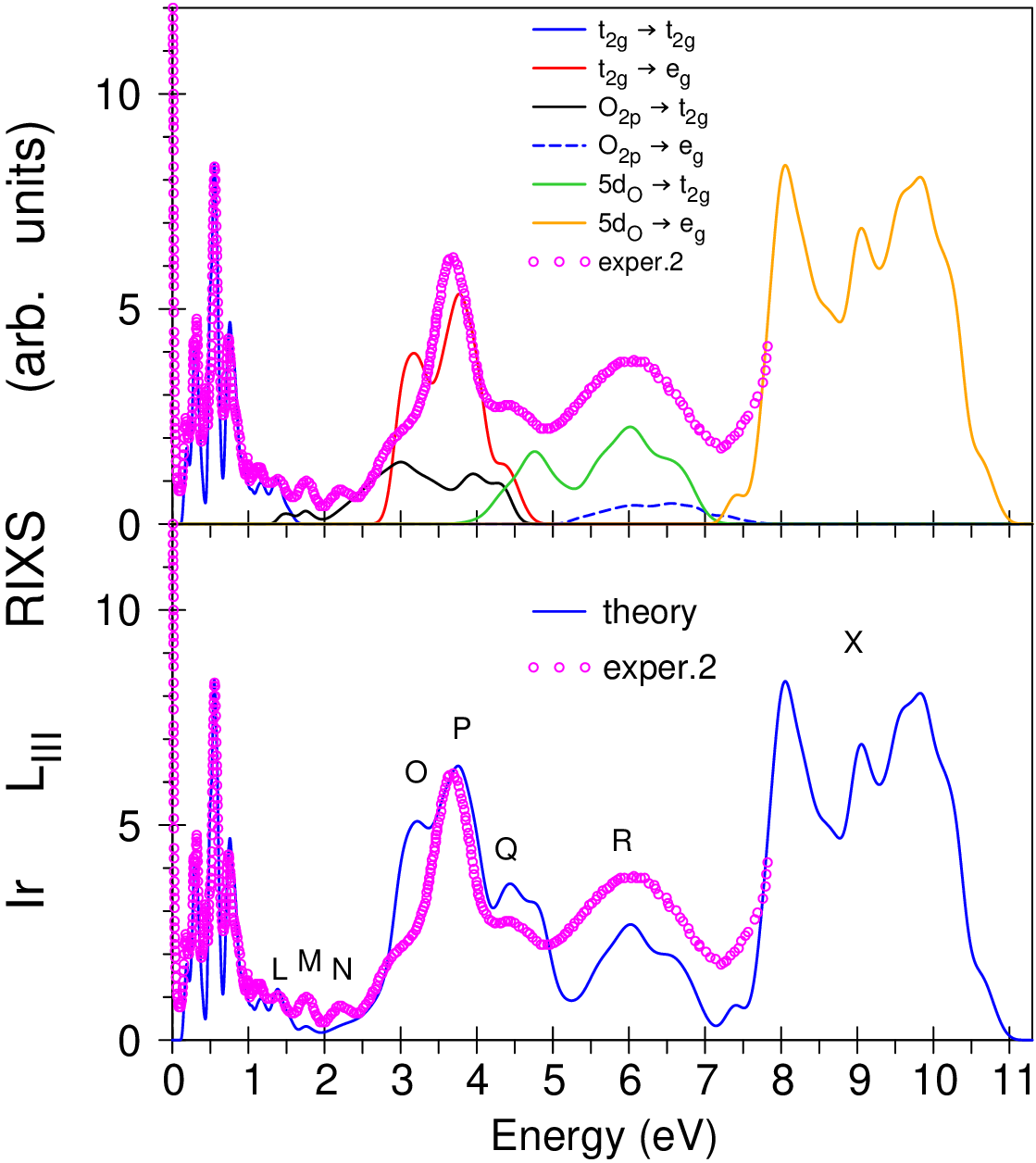}
\end{center}
\caption{\label{rixs_BAIO}(Color online) Lower panel: the experimental
  RIXS spectrum (open magenta
  circles) measured by Katukuri {\it et al.}  \cite{KLM+22} for the
  momentum transfer vector {\bf Q} = (23.5, 0, 2.5) in reciprocal
  lattice units and incident photon energy $E_i$ = 11215 eV at the Ir
  $L_3$ edge in Ba$_5$AlIr$_2$O$_{11}$ compared with the theoretically
  calculated one in the GGA+SO+$U$ approach ($U_{\rm{eff}}$=0.6 eV).
  Upper panel: the experimental RIXS spectrum (open
  magenta circles) measured by Katukuri {\it et al.} \cite{KLM+22} at
  the Ir $L_3$ edge and the decomposition of the Ir $L_3$ RIXS
  spectrum into different interband transitions. }
\end{figure}

Figure \ref{rixs_BAIO} (the lower panel) shows the experimental RIXS spectrum
(open magenta circles) measured by Katukuri {\it et al.}  \cite{KLM+22} at the
Ir $L_3$ edge in Ba$_5$AlIr$_2$O$_{11}$ in a wide energy interval up to 8 eV
compared with the theoretically calculated one in the GGA+SO+$U$ approach.  As
we mentioned above, the interband transitions below 1.5 eV (the blue curve in
the upper panel of Fig. \ref{rixs_BAIO}) are due to the {\tg} $\rightarrow$
{\tg} transitions. The intensive peak $P$ at $\sim$3.5 eV with the low energy
shoulder $O$ is mostly due to $\tg \rightarrow \eg$ transitions (the red curve
in the upper panel of Fig. \ref{rixs_BAIO}). The O$_{2p}$ $\rightarrow \tg$
transitions (the black curve in the upper panel of Fig. \ref{rixs_BAIO}) also
contribute to this peak and the low energy shoulder $O$ as well as to small
peaks $M$ and $N$. The next fine structure $R$ from 5 to 7 eV is due to
5$d_{\rm{O}}$ $\rightarrow$ {\tg} transitions (the green curve). The shoulder
$Q$ is partly due to the 5$d_{\rm{O}}$ $\rightarrow$ {\tg} transitions as well
as due to the $\tg \rightarrow \eg$ transitions. The O$_{2p}$ $\rightarrow$
{\eg} transitions are very weak (the dashed blue curve). The high energy
intensive peak $X$ between 7.5 and 11 eV is due to 5$d_{\rm{O}}$ $\rightarrow$
{\eg} interband transitions.

\begin{figure}[tbp!]
\begin{center}
\includegraphics[width=0.9\columnwidth]{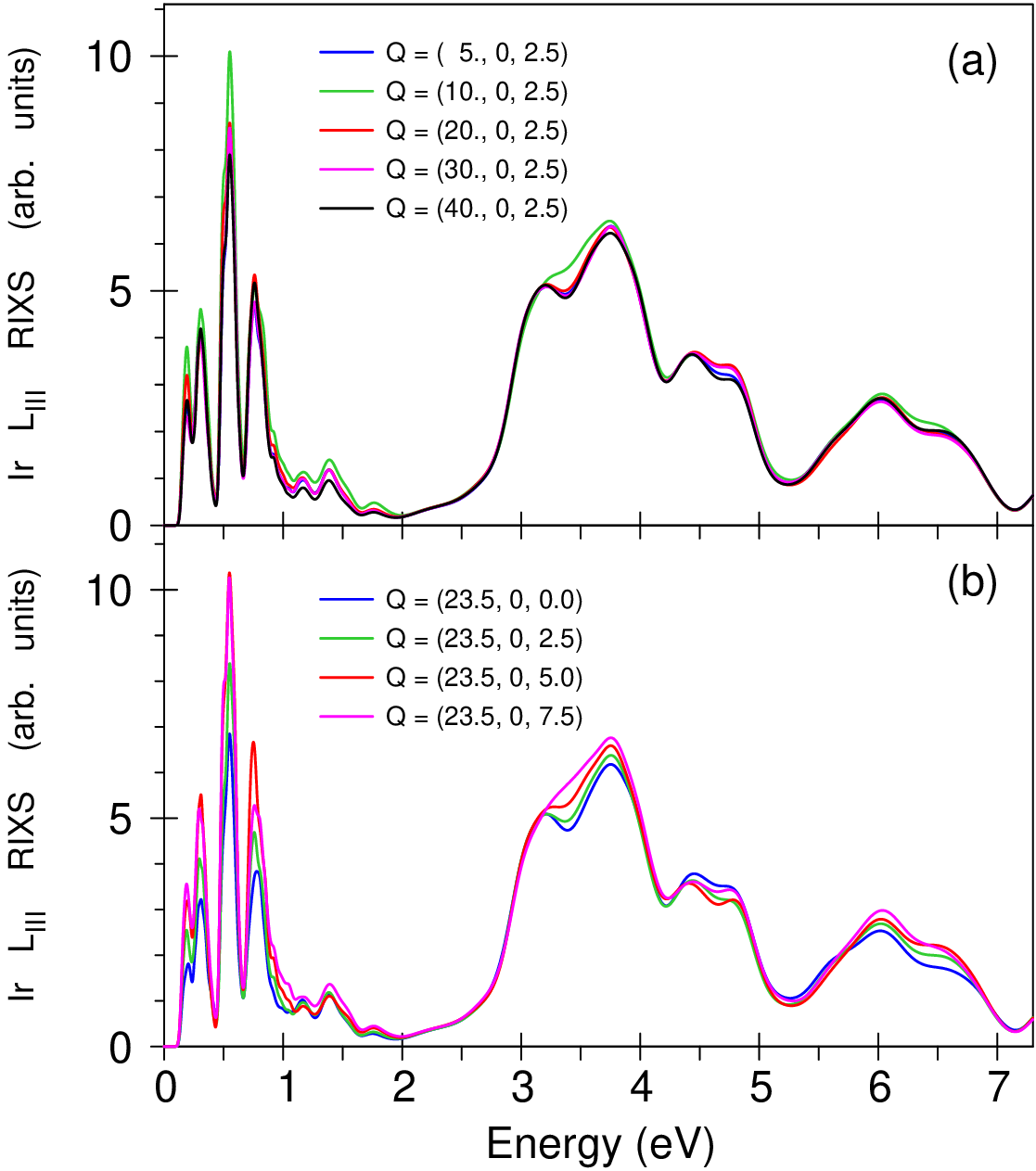}
\end{center}
\caption{\label{rixs_Ir_L3_Q_BAIO}(Color online) The RIXS
  spectra at the Ir $L_3$ edge in
  Ba$_5$AlIr$_2$O$_{11}$ calculated as a function of $Q_x$ (a) and
  $Q_z$ (b) with the momentum transfer vector {\bf Q} = (Q$_x$, 0,
  Q$_z$) in reciprocal lattice units for incident photon energy
  $\hbar \omega_{in}$ = 11215 eV. }
\end{figure}

It is widely believed that $d-d$ excitations show only small momentum transfer
vector {\bf Q} dependence in 5$d$ transition metal compounds.  Figure
\ref{rixs_Ir_L3_Q_BAIO}(a) shows the RIXS spectra at the Ir $L_3$ edge in
Ba$_5$AlIr$_2$O$_{11}$ calculated as a function of $Q_x$ with the momentum
transfer vector {\bf Q} = (Q$_x$, 0, 2.5) for incident photon energy $\hbar
\omega_{in}$ = 11215 eV. We found that the RIXS spectra are quite similar for
Q$_x$ changing from 5 to 40 in reciprocal lattice units. With increasing Q$_z$
for the momentum transfer vector {\bf Q} = (23.5, 0, Q$_z$), the low energy
peaks below 1.5 eV are slightly increased while the high energy fine
structures between 2 and 7 eV are changed insignificantly
[Fig. \ref{rixs_Ir_L3_Q_BAIO}(b)].

Analyzing Fig. \ref{rixs_Ir_L3_Q_BAIO} we can conclude that the momentum
dependence of the excitations in Ba$_5$AlIr$_2$O$_{11}$ is rather small, as it
was earlier observed in other iridates such as Sr$_3$CuIrO$_6$ \cite{LKH+12},
In$_2$Ir$_2$O$_7$ \cite{KTD+20}, or Sr$_2$IrO$_4$ \cite{AKB24}.

\begin{figure}[tbp!]
\begin{center}
\includegraphics[width=0.9\columnwidth]{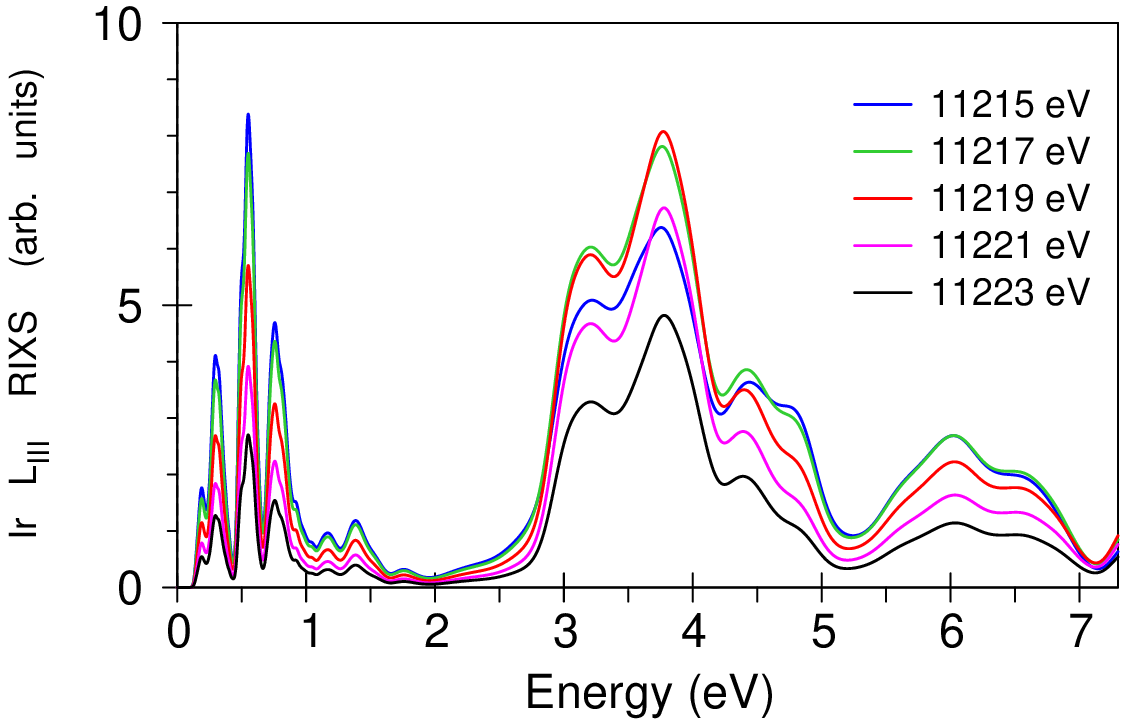}
\end{center}
\caption{\label{rixs_Ir_L3_Ei_BAIO}(Color online) The RIXS
  spectra as a function of incident
  photon energy $E_i$ calculated at the Ir $L_3$ edge in
  Ba$_5$AlIr$_2$O$_{11}$ with the momentum transfer vector {\bf Q} =
  (23.5, 0, 2.5) in reciprocal lattice units calculated in the
  GGA+SO+$U$ approach ($U_{\rm{eff}}$=0.6 eV). }
\end{figure}

Figure \ref{rixs_Ir_L3_Ei_BAIO} shows the Ir $L_3$ RIXS spectrum as a function
of incident photon energy $E_i$ above the corresponding edge with the momentum
transfer vector {\bf Q} = (23.5, 0, 2.5). We found that the low energy fine
structure $\le$1.5 eV corresponding to intra-{\tg} excitations is steadily
decreased when the incident photon energy changes from 11215 to 11223 eV,
whereas the high energy peak corresponding to the $\tg \rightarrow \eg$
transitions shows more complex behavior. The intensity of the peak is
increased with $E_i$ varying from 11215 to 11219 eV, but then the peak is
decreased with $E_i$ increasing to 11221 and 11223 eV.

\subsection{RIXS spectra at the Ir $K$ and M$_5$ edges}

\begin{figure}[tbp!]
\begin{center}
\includegraphics[width=0.9\columnwidth]{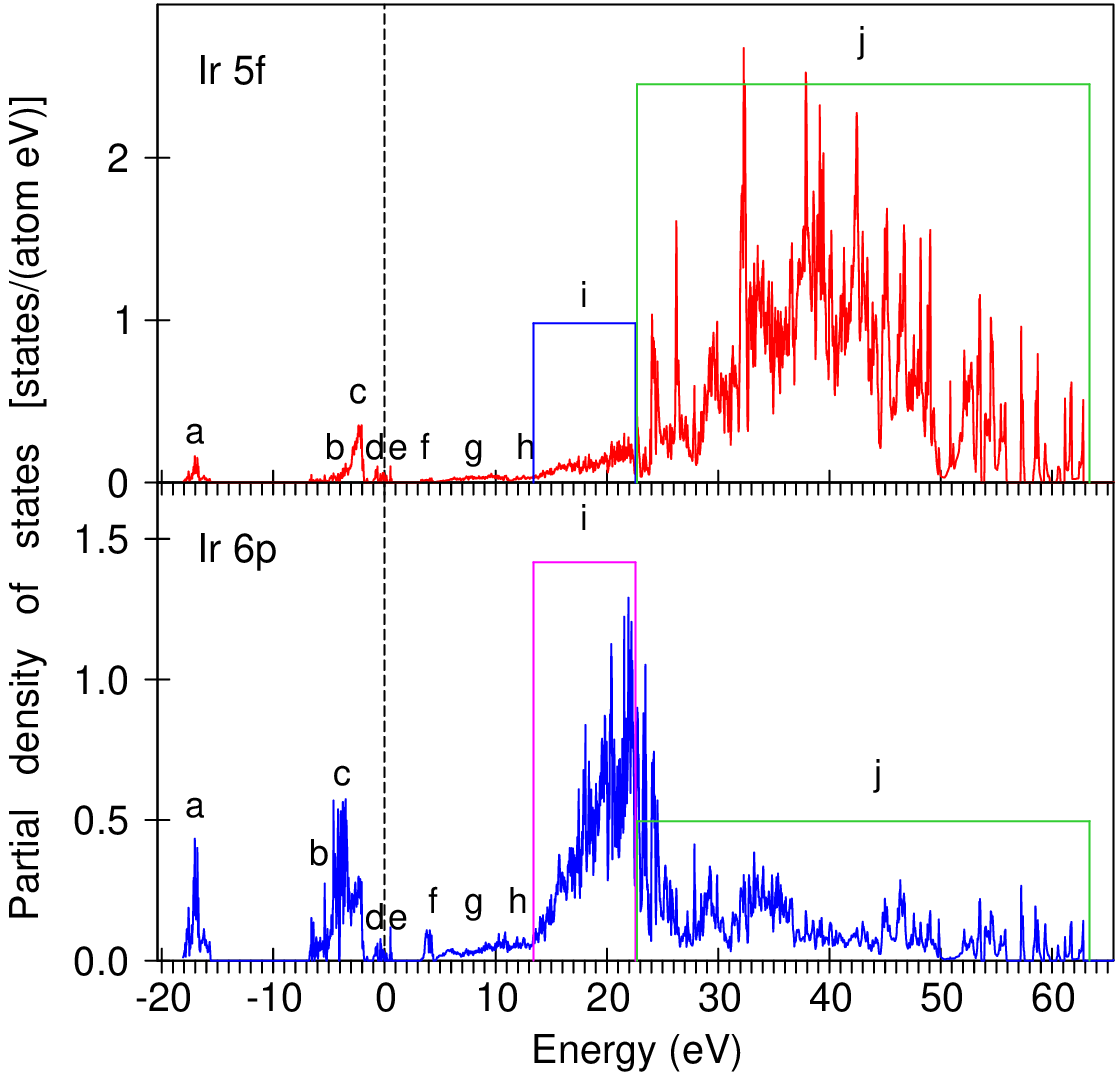}
\end{center}
\caption{\label{PDOS_Ir_6p_5f_BAIO}(Color online) The Ir 6$p$ (the
  lower panel) and 5$f$ (the upper panel) partial DOS [in states/(atom
  eV)] in Ba$_5$AlIr$_2$O$_{11}$ calculated in the GGA+SO+$U$
  ($U_{\rm{eff}}$= 0.6 eV) approach. }
\end{figure}

Let us consider now the RIXS spectra at the Ir $K$ and $M_5$ edges.  For that
we first present in Fig. \ref{PDOS_Ir_6p_5f_BAIO} the Ir 6$p$ (the lower
panel) and 5$f$ (the upper panel) partial DOS in a wide energy interval from
$-$21 to 62 eV. We distinguish several groups of bands. The group $a$ derives
from the hybridizations of Ir 6$p$ and 5$f$ states with oxygen 2$s$
states. The groups $b$ and $c$ are due to the hybridization of these states
with oxygen 2$p$ states. The groups $d$ and $e$ are from the hybridization
with Ir {\tg} LEB and UEB, respectively. The group $f$ comes from the
hybridization with Ir {\eg} states. The group $g$ comes from the hybridization
with Ba 5$d$ and 4$f$ states, and the group $h$ is due to the hybridization
with Al 3$p$ states. The structure $i$ is the Ir 6$p$ band itself. The
structure $j$ is the Ir 5$f$ band.

\begin{figure}[tbp!]
\begin{center}
\includegraphics[width=0.9\columnwidth]{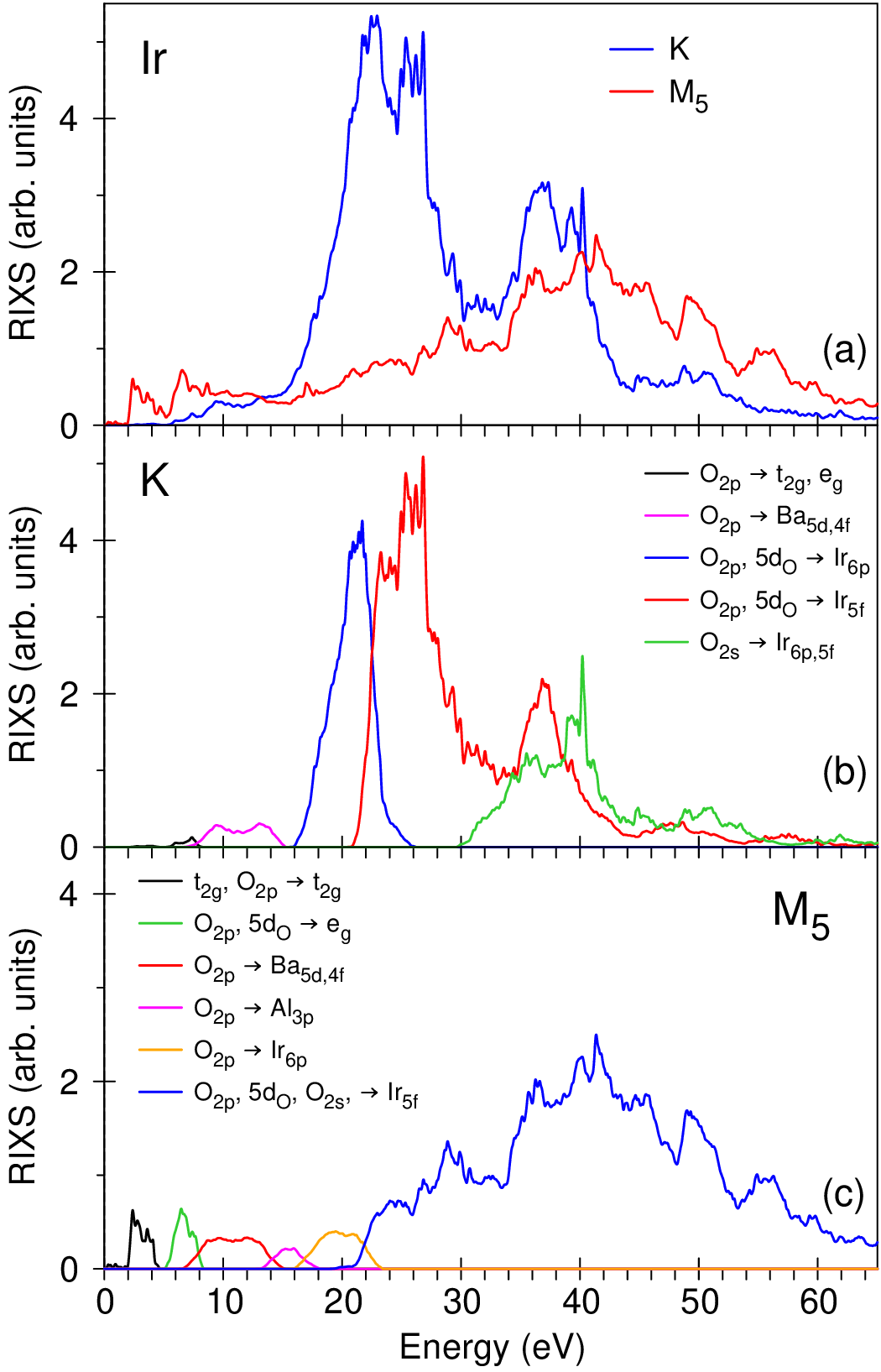}
\end{center}
\caption{\label{rixs_Ir_K_M5_BAIO}(Color online) (a) The theoretically
  calculated RIXS spectrum at
  the Ir $K$ (the blue curve) and $M_5$ (the red curve) edges in
  Ba$_5$AlIr$_2$O$_{11}$; (b) the decomposition of the Ir $K$ RIXS
  spectrum into different interband transitions; (c) the decomposition
  of the Ir $M_5$ RIXS spectrum into different interband
  transitions. }
\end{figure}

Figure \ref{rixs_Ir_K_M5_BAIO}(a) presents the theoretically calculated Ir $K$
(the blue curve) and Ir $M_5$ (the red curve) RIXS spectra in
Ba$_5$AlIr$_2$O$_{11}$. The spectra significantly differ from each other and
from the RIXS spectra at the Ir $L_3$ edge. The partial contributions from
different interband transitions are presented in
Figs. \ref{rixs_Ir_K_M5_BAIO}(b) and \ref{rixs_Ir_K_M5_BAIO}(c) for the Ir $K$
and $M_5$ edges, respectively. Due to significantly smaller spatial
localization of the Ir 5$f$ orbitals in comparison with the 6$p$ ones the
peaks $a$, $b$, $c$, $d$, and $e$, which originate from the hybridization with
oxygen 2$s$ and 2$p$, and Ir {\tg} and {\eg} states, respectively, possess
much smaller intensity in the Ir 5$f$ partial DOS in comparison with the Ir
6$p$ one (Fig. \ref{PDOS_Ir_6p_5f_BAIO}). However, the low energy part of the
Ir $M_5$ spectrum $\le$10 eV possesses larger intensity in comparison with the
Ir $K$ RIXS spectrum due to the corresponding matrix elements. This part of
the $M_5$ spectrum is due to ({\tg}, O$_{2p}$, Ir$_{5d_O}$) $\rightarrow$
({\tg}, {\eg}) transitions. The major contribution to the Ir $M_5$ RIXS
spectrum above 20 eV is due to the interband transitions into Ir 5$f$ states
[the blue curve in Fig. \ref{rixs_Ir_K_M5_BAIO}(c)].

The Ir $K$ RIXS spectrum consists of two major peaks at 8-30 and 32-44 eV. The
low energy part of the first peak is due to the interband transitions from
O$_{2p}$ and 5$d_O$ states into empty Ir$_{6p}$ states [the blue curve in
  Fig. \ref{rixs_Ir_K_M5_BAIO}(b)]. The transitions (O$_{2p}$, 5$d_O$)
$\rightarrow$ Ir$_{5f}$ contribute to both peaks [the red curve in
  Fig. \ref{rixs_Ir_K_M5_BAIO}(b)]. We can also see a visible contribution
into the second peak at 32-44 eV of O$_{2s}$ $\rightarrow$ Ir$_{6p,5f}$
transitions (the green curve). Experimental measurements of the RIXS spectra
at the Ir $K$ and $M_5$ edges are highly desirable.

\subsection{RIXS spectra at the Ba $K$, $L_3$, and M$_5$ edges}

\begin{figure}[tbp!]
\begin{center}
\includegraphics[width=0.9\columnwidth]{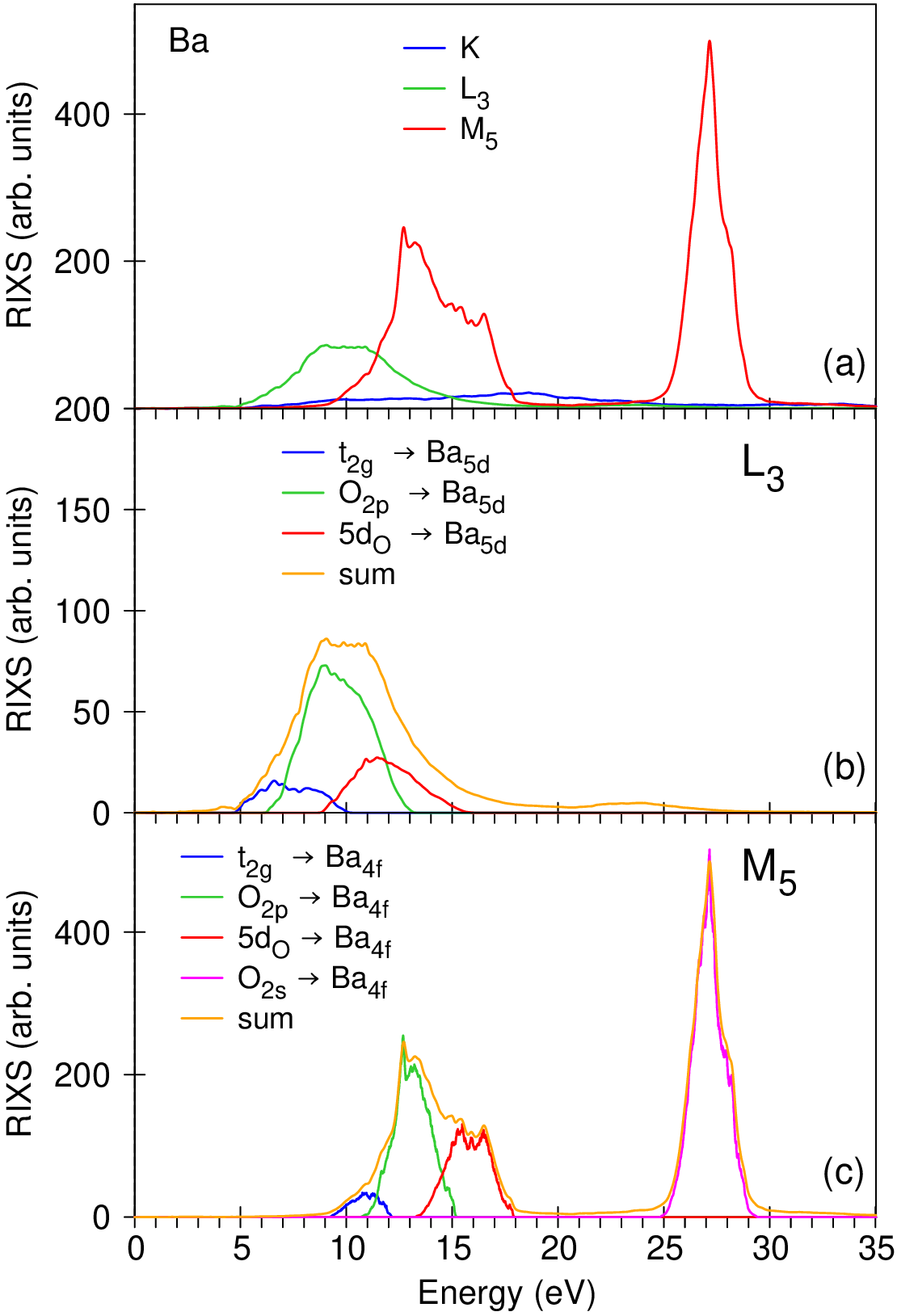}
\end{center}
\caption{\label{rixs_Ba_BAIO}(Color online) (a) The theoretically
  calculated RIXS spectrum at
  the Ba $K$ (the blue curve), $L_3$ (the green curve), and $M_5$ (the red curve)
  edges in Ba$_5$AlIr$_2$O$_{11}$; (b) the decomposition of the Ba
  $L_3$ RIXS spectrum into different interband transitions; (c) the
  decomposition of the Ba $M_5$ RIXS spectrum into different interband
  transitions. }
\end{figure}

Figure \ref{rixs_Ba_BAIO}(a) presents the theoretically calculated Ba $K$ (the
blue curve), $L_3$ (the green curve) and $M_5$ (the red curve) RIXS spectra in
Ba$_5$AlIr$_2$O$_{11}$. These spectra significantly differ from each other and
from the RIXS spectra at the Ir $K$, $L_3$, and $M_5$ edges. The partial
contributions from different interband transitions are presented in
Figs. \ref{rixs_Ba_BAIO}(b) and \ref{rixs_Ba_BAIO}(c) for the Ba $L_3$ and
$M_5$ edges, respectively.

The Ba $M_5$ RIXS spectrum is the most intensive and consists of two peaks
separated by a rather wide energy gap: a low energy peak between 9 and 18 eV
and a high energy narrow peak at 25$-$29 eV. The latter peak is completely due
to O$_{2s}$ $\rightarrow$ Ba$_{4f}$ transitions [the magenta curve in Fig.
  \ref{rixs_Ba_BAIO}(c)]. The low energy peak is devoted to the interband
transitions from {\tg}, O$_{2p}$, and 5$d_O$ states into Ba$_{4f}$ ones. The
Ba $L_3$ RIXS spectrum possesses a single peak between 5 and 16 eV, which is
formed by the transitions from the {\tg}, O$_{2p}$, and 5$d_O$ states into Ba
5$d$ states [see Fig. \ref{rixs_Ba_BAIO}(b)]. The Ba $K$ RIXS spectrum
possesses relatively small intensity and is formed by the interband
transitions from O$_{2s}$, O$_{2p}$, and 5$d_O$ states into Ba 5$d$, 4$f$, Al
3$p$, and Ir 6$p$ states with almost equal intensity (not shown). Experimental
measurements of the RIXS spectra at the Ba $K$, $L_3$ and $M_5$ edges could be
quite perspective and are highly desirable.

\subsection{XAS and RIXS spectra at the O $K$ edge}
\label{sec:rixs_O}

In the XAS, XMCD, and RIXS processes at the O $K$ edge, the 1$s$ core level is
involved. The exchange splitting of the 1$s$-core state is extremely small and
SOC is absent for O 1$s$ orbitals, therefore, only the exchange and spin-orbit
splitting of the 2$p$ states is responsible for the observed spectra at the
oxygen $K$ edge. However, the oxygen valence 2$p$ states because of their
delocalized nature are sensitive to the electronic states at neighboring 5$d$
sites. They strongly hybridize with the 5$d$ orbitals. Due to such
hybridization combined with high SOC at the 5$d$ ion, information on the
elementary excitations can be extracted using an indirect RIXS process at the
O $K$ edge \cite{LOH+18}. Although O $K$ RIXS has a much smaller penetration
depth ($\sim$100 nm) than 5$d$ $L$ RIXS, a comparison between O $K$ and Ir
$L_3$ spectra measured, for example, for Sr$_2$IrO$_4$ suggests that they have
comparable counting efficiency \cite{LOH+18}. The lower penetration depth of
soft x-rays has its own advantages providing high sensitivity to ultrathin
samples such as films. Soft x-ray RIXS at the O $K$ edge is a promising method
for studying the electronic and magnetic excitations in 5$d$ compounds. The
RIXS spectra as well as XAS spectra at the oxygen $K$ edge in
Ba$_5$AlIr$_2$O$_{11}$ were investigated experimentally by Katukuri {\it et
  al.} \cite{KLM+22} for different polarizations.

\begin{figure}[tbp!]
\begin{center}
\includegraphics[width=0.9\columnwidth]{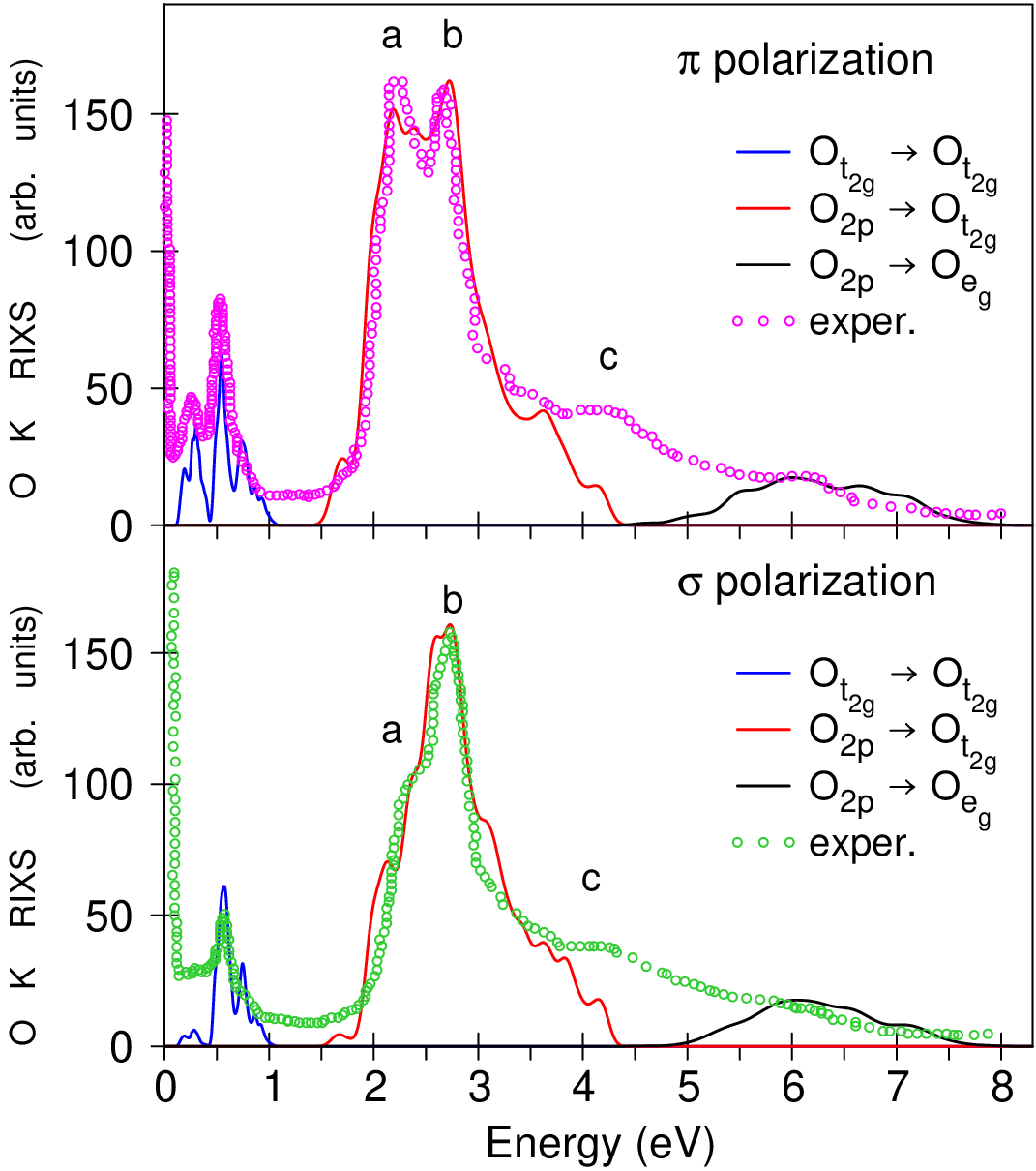}
\end{center}
\caption{\label{rixs_O_BAIO}(Color online) The RIXS spectra in
  Ba$_5$AlIr$_2$O$_{11}$ measured by Katukuri {\it et al.}
  \cite{KLM+22} at the O $K$ edge for grazing incidence and excitation
  energy $E_i$ = 527.6 eV with the momentum transfer vector {\bf Q} =
  (0.93, 0, 0.65) in reciprocal lattice units for the $\sigma$ (open
  green circles in the lower panel) and $\pi$ (open magenta circles in
  the upper panel) polarizations in comparison with the spectra calculated in the
  GGA+SO+$U$ ($U_{\rm{eff}}$= 0.6 eV) approach. }
\end{figure}

Figure \ref{rixs_O_BAIO} presents the RIXS spectra in Ba$_5$AlIr$_2$O$_{11}$
measured by Katukuri {\it et al.} \cite{KLM+22} at the O $K$ edge for grazing
incidence and excitation energy $E_i$ = 527.6 eV with the momentum transfer
vector {\bf Q} = (0.93, 0, 0.65) for the $\sigma$ (open green circles in the
lower panel) and $\pi$ (open magenta circles in the upper panel) polarizations
in comparison with the spectra calculated in the GGA+SO+$U$ approach. The O
$K$ RIXS spectra consist of the elastic peak centered at zero energy loss and
three major inelastic excitations $\le$1 eV, 1.8$-$4.7 eV, and $>$5 eV. We
found that the low energy features $\le$1 eV (the blue curves in
Fig. \ref{rixs_O_BAIO}) are due to the interband transitions between occupied
and empty O$_{\tg}$ states, which appear as a result of the strong
hybridization between oxygen 2$p$ states with Ir {\tg} LEB and UEB in close
vicinity to the Fermi level (Fig. \ref{PDOS_BAIO}). The next major fine
structures between 1.8 and 4.7 eV (the red curves in Fig. \ref{rixs_O_BAIO})
consist of two peaks $a$ and $b$ and a high energy shoulder $c$. These fine
structures reflect the interband transitions from occupied O 2$p$ states to
empty oxygen states, which originate from the hybridization with Ir {\tg}
states. The peak $a$ is significantly suppressed for the $\sigma$ polarization
in comparison with the $\pi$ one. The theory reproduces well the shapes and
energy positions and relative intensity of the low energy features $a$ and $b$
for both polarizations, but not the high energy shoulder $c$, which is
underestimated in our calculations.  The O$_{2p}$ $\rightarrow$ {\eg}
transitions situated between 5 and 8 eV energy interval (the black curves in
Fig. \ref{rixs_O_BAIO}) have quite small intensity and appear in the
experimental spectrum just as a tail.

\begin{figure}[tbp!]
\begin{center}
\includegraphics[width=0.9\columnwidth]{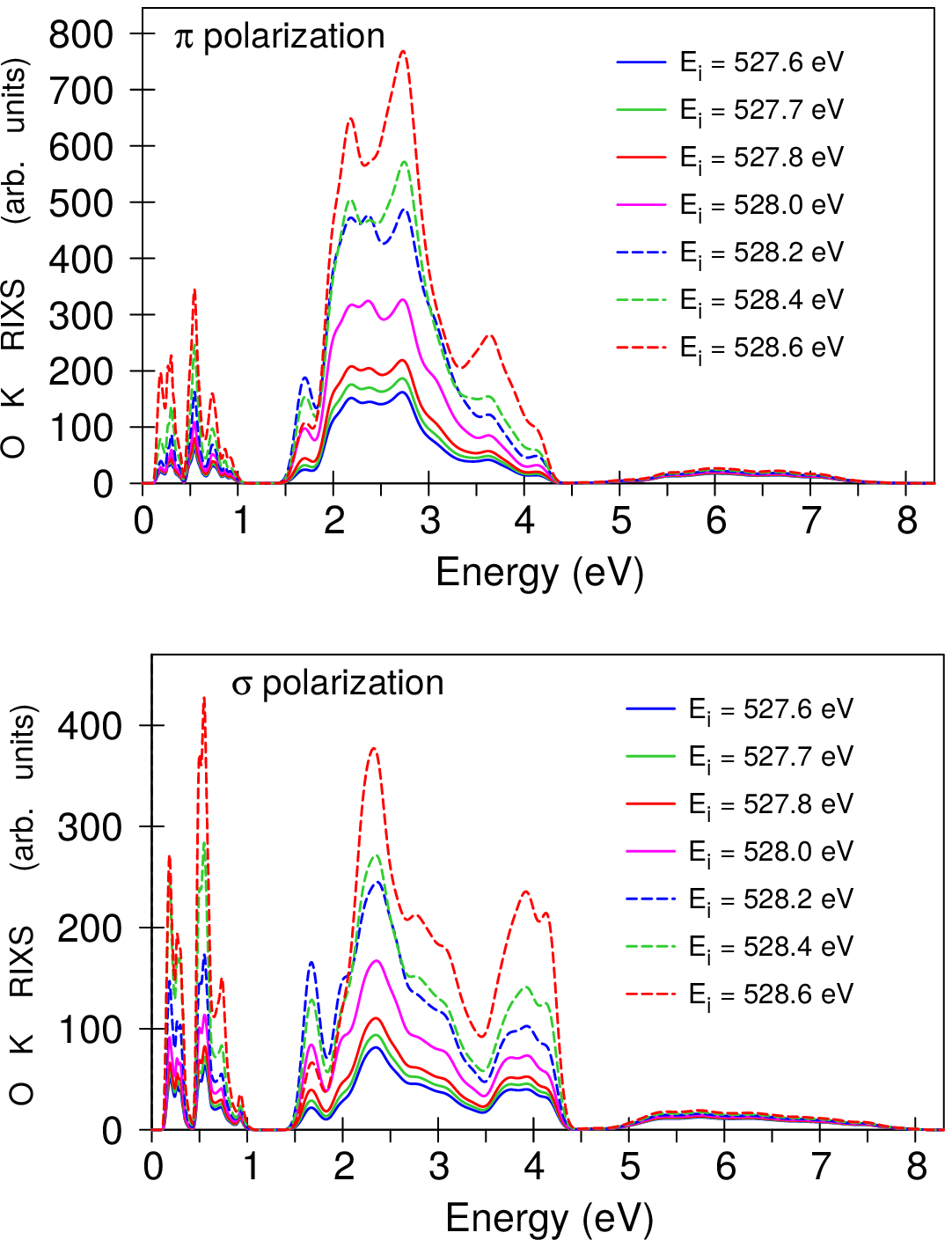}
\end{center}
\caption{\label{rixs_O_Ei_BAIO}(Color online) The RIXS
  spectra at the O $K$ edge as a function of incident photon
  energy for the $\sigma$ (the lower panel) and $\pi$ (the upper panel)
  polarizations calculated in the GGA+SO+$U$ ($U_{\rm{eff}}$=0.6 eV)
  approach. }
\end{figure}

Figure \ref{rixs_O_Ei_BAIO} presents the RIXS spectra as a function of
incident photon energy $E_i$ calculated at the O $K$ edge in
Ba$_5$AlIr$_2$O$_{11}$ for two polarizations. We found much stronger
dependence on the incident photon energy in the case of the O $K$ RIXS
spectrum in comparison with the corresponding dependence at the Ir $L_3$ edge
(compare Figs. \ref {rixs_Ir_L3_Ei_BAIO} and \ref{rixs_O_Ei_BAIO}). With
increasing the incident photon energy both the peaks $\le$1 eV and the peaks
between 1.8 and 4.7 eV are significantly increased for both
polarizations. This occurs in a small energy interval of 1 eV for $E_i$ from
527.6 to 528.6 eV. However, the O$_{2p}$ $\rightarrow$ {\eg} interband
transitions, situated in the 5$-$8 eV energy interval, possess quite small
incident photon energy dependence for both polarizations.

\begin{figure}[tbp!]
\begin{center}
\includegraphics[width=0.9\columnwidth]{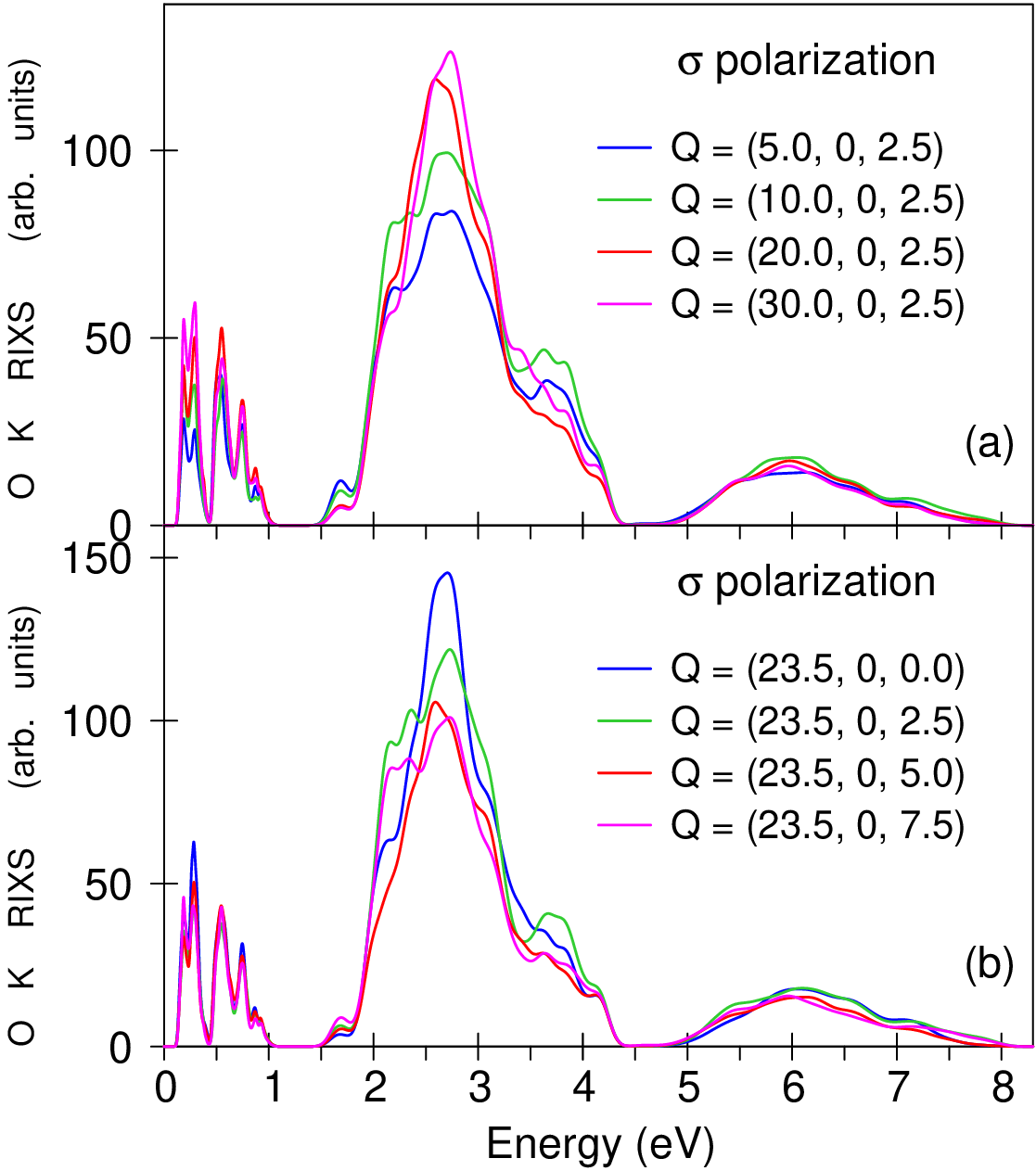}
\end{center}
\caption{\label{rixs_O_Q_BAIO}(Color online) The RIXS
  spectra at the O $K$ edge in
  Ba$_5$AlIr$_2$O$_{11}$ calculated as a function of $Q_x$ (a) and
  $Q_z$ (b) with the momentum transfer vector {\bf Q} = (Q$_x$, 0,
  Q$_z$) in reciprocal lattice units for incident photon energy
  $\hbar \omega_{in}$ = 527.6 eV. }
\end{figure}

Figure \ref{rixs_O_Q_BAIO}(a) shows the RIXS spectra at the O $K$ edge in
Ba$_5$AlIr$_2$O$_{11}$ calculated as a function of $Q_x$ with the momentum
transfer vector {\bf Q} = (Q$_x$, 0, 2.5) for incident photon energy $\hbar
\omega_{in}$ = 527.6 eV and the $\sigma$ polarization. We found that the {\tg}
$\rightarrow$ {\tg} transitions at $\le$1 eV energy interval as well as the
O$_{2p}$ $\rightarrow$ {\tg} transitions between 1.8 and 4.5 eV are
monotonously increased for Q$_x$ changing from 5 to 30 in reciprocal lattice
units, while the O$_{2p}$ $\rightarrow$ {\eg} transitions between 5 and 8 eV
are changed insignificantly. With increasing Q$_z$, the low energy peaks
$\le$1 eV are changed insignificantly [for the momentum transfer vector {\bf
    Q} = (23.5, 0, Q$_z$)] and the high energy fine structures between 1.8 and
4.5 eV are decreased [Fig. \ref{rixs_O_Q_BAIO}(b)]. The O$_{2p}$ $\rightarrow$
{\eg} transitions are less sensitive to the momentum transfer vector {\bf Q}.

\begin{figure}[tbp!]
\begin{center}
\includegraphics[width=0.9\columnwidth]{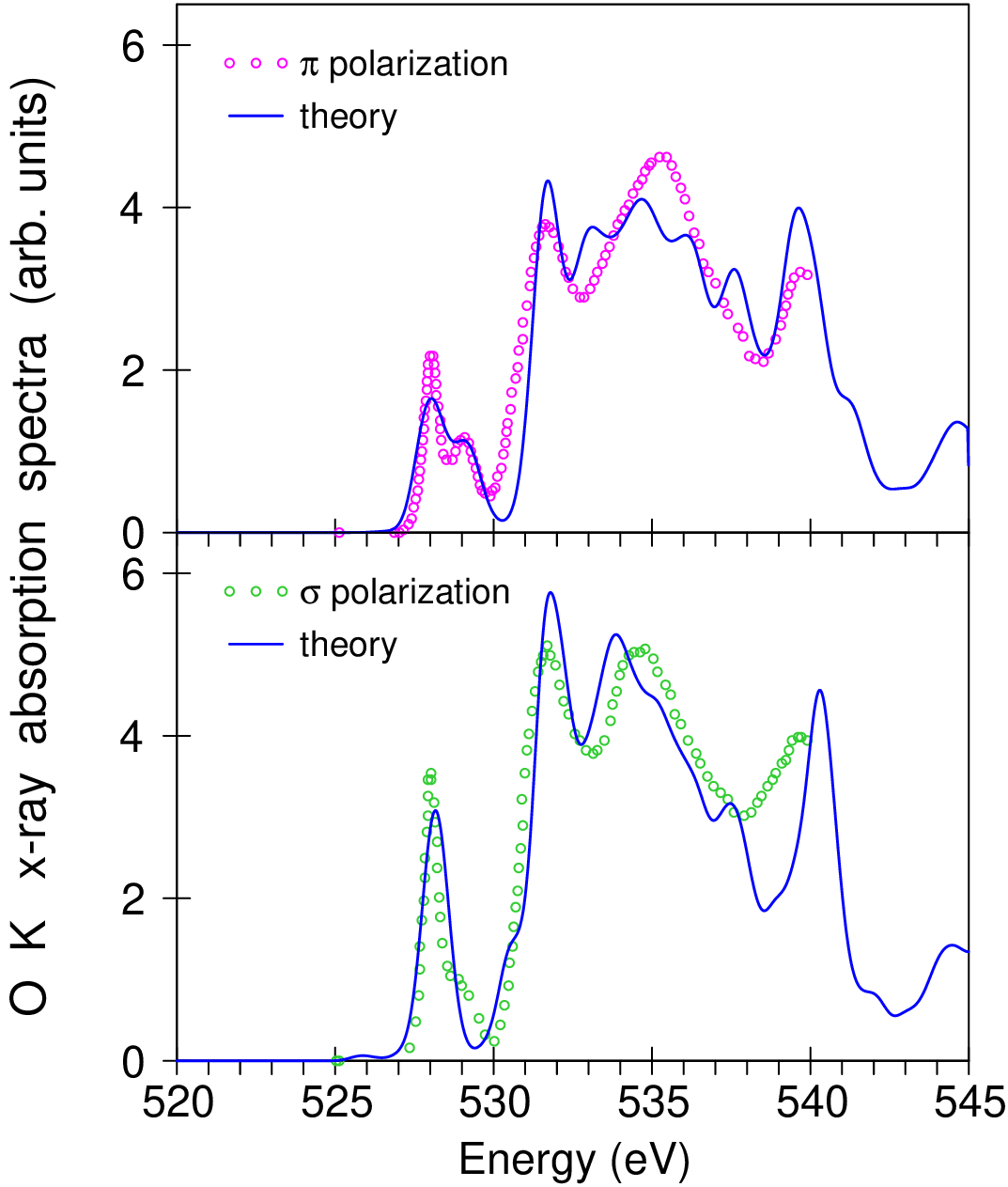}
\end{center}
\caption{\label{XAS_O_BAIO}(Color online) The experimental O $K$
  polarization-dependent XAS spectra
  (open circles) \cite{KLM+22} in Ba$_5$AlIr$_2$O$_{11}$ compared with
  the theoretically calculated ones in the GGA+SO+$U$ approach
  ($U_{\rm{eff}}$ = 0.6 eV). }
\end{figure}

Figure \ref{XAS_O_BAIO} presents the experimental O $K$ XAS spectra for the
$\pi$ (open magenta circles in the upper panel) and $\sigma$ polarizations
(open green circles in the lower panel) \cite{KLM+22} compared with the
theoretically calculated ones in the GGA+SO+$U$ approach. The theory
reproduces quite well the polarization dependence of the O $K$ XAS
spectrum. There is a single quite intensive low energy peak at 528 eV for the
$\sigma$ polarization. The corresponding peak for the $\pi$ polarization is
split into two peaks with relatively smaller intensity.  Although the theory
reproduces relatively well the shape of the peak at 530-538 eV for the
$\sigma$ polarization, for the $\pi$ polarization our calculations show
several peaks at 533-539 eV instead of one as in the experiment.

\section{Conclusions}

To summarize, we have investigated the electronic and magnetic structures of
Ba$_5$AlIr$_2$O$_{11}$ in the frame of the fully relativistic spin-polarized
Dirac approximation. We also present comprehensive theoretical calculations of
the RIXS spectra at the Ir $K$, $L_3$, $M_5$, Ba $K$, $L_3$, $M_5$ and oxygen
$K$ edges, the XAS and XMCD spectra at the Ir $L_{2,3}$ edges, and polarizaion
dependence of the O $K$ XAS spectra.

Ba$_5$AlIr$_2$O$_{11}$ is a Mott insulator that undergoes a subtle structural
phase transition near $T_S$ = 210 K and a transition to FiM order at $T_M$ =
4.5 K. The GGA+SO band structure calculations produce a metallic solution for
Ba$_5$AlIr$_2$O$_{11}$ in contradiction to experiment. To produce the correct
dielectric ground state one has to take into account strong Coulomb
correlations in Ba$_5$AlIr$_2$O$_{11}$. The crystal structure of
Ba$_5$AlIr$_2$O$_{11}$ consists of IrO$_6$ octahedra sharing a face along the
crystallographic $b$ axis, and develop so called Ir$_2$O$_9$ dimers. Each
dimer consists of two inequivalent octahedral Ir$_1$ and Ir$_2$ sites. It was
supposed that Ba$_5$AlIr$_2$O$_{11}$ possesses charge order with Ir$_1$ and
Ir$_2$ sites occupied by pentavalent Ir$^{5+}$ (5$d^4$) and tetravalent
Ir$^{4+}$ (5$d^5$) ions, respectively. However, we have found that the charge
disproportionation of $\sim$0.3 electron is not complete. A purely ionic model
with strong SOC, which would support $J_{\rm{eff}}$ = $\frac{1}{2}$ in the
Ir$_2^{4+}$ (5$d^5$) ions and $J_{\rm{eff}}$ = 0 in the Ir$_1^{5+}$ (5$d^4$)
ions, is not entirely applicable in Ba$_5$AlIr$_2$O$_{11}$. Our GGA+SO+$U$
calculations produce the ionicity equal to +4.3 and +4.7 for Ir$_2$ and
Ir$_1$, respectively.

The remarkably large branching ratio BR = $I_{L_3}/I_{L_2}$ = 4.09 and 3.70
for the Ir$_2$ and Ir$_1$ sites, respectively, indicates strong SO effects in
Ba$_5$AlIr$_2$O$_{11}$. There is a relatively large XMCD signal at the $L_3$
edge for the Ir$_2$ site. However, for the Ir$_1$ site the dichroism is much
smaller due to the smallness of the orbital magnetic moment at that site.

The theoretically calculated Ir $L_3$ RIXS spectrum is in good agreement with
the experiment. We have found that the low energy peaks $\le$1.5 eV correspond
to intra-{\tg} excitations. There are five separated groups of {\tg} bands
below $E_F$ and three groups of empty bands, which produce three narrow DOS
peaks separated by energy gaps. The interband transitions between these five
occupied and three empty groups of bands produce quite a rich fine structure
of the Ir $L_3$ RIXS spectrum consisting of twelve well distinguished peaks
below $\le$1.5 eV. The appearance of multiple peaks in the RIXS spectrum is a
direct consequence of the strong noncubic crystal field splitting originating
from the distorted octahedral environment of the Ir ions. The intensive peak
at $\sim$3.5 eV is mostly due to $\tg \rightarrow \eg$ transitions. The next
fine structure from 4 to 7 eV is mostly due to 5$d_{\rm{O}}$ $\rightarrow$
{\tg} transitions. The high energy intensive peak between 7.5 and 11 eV is due
to 5$d_{\rm{O}}$ $\rightarrow$ {\eg} interband transitions.

We have found that the momentum dependence of the excitations in
Ba$_5$AlIr$_2$O$_{11}$ is rather small, as it was earlier observed in other
iridates such as Sr$_3$CuIrO$_6$, In$_2$Ir$_2$O$_7$, or Sr$_2$IrO$_4$. We have
also investigated the Ir $L_3$ RIXS spectrum as a function of incident photon
energy $E_i$ and found that the low energy fine structure corresponding to
intra-{\tg} excitations is slightly decreased when the incident photon energy
changes from 11210 to 11212 eV, whereas the high energy peak corresponding to
the $\tg \rightarrow \eg$ transitions is monotonically increased.

The RIXS spectra at the Ir $K$ and $M_5$ edges occupy quite a wide energy
interval up to 60 eV, which almost six times larger than, for example, the
occupation interval of the Ir $L_3$ or oxygen $K$ spectra. The major
contribution to the Ir $M_5$ RIXS spectrum comes from the interband
transitions into empty Ir 5$f$ bands. The Ir $K$ spectrum reflects the energy
distribution of different states in Ba$_5$AlIr$_2$O$_{11}$ (oxygen 2$s$ and
2$p$, Ir 5$d$ and 5$f$) due to the extended character of Ir 6$p$
orbitals. Experimental measurements of the RIXS spectra at the Ir $K$ and
$M_5$ edges could be quite perspective.

Another useful possibility is the extention of the RIXS measurements on the Ba
$K$, $L_3$, and M$_5$ edges. These spectra also significantly differ from each
other and from the RIXS spectra at the Ir $K$, $L_3$, and $M_5$ edges. The Ba
$M_5$ RIXS spectrum is the most intensive and consists of two peaks separated
by a rather wide energy gap: a low energy peak between 9 and 18 eV and a high
energy narrow peak at 25$-$29 eV. The latter peak is completely due to
O$_{2s}$ $\rightarrow$ Ba$_{4f}$ transitions. The low energy peak is devoted
to the interband transitions from {\tg}, O$_{2p}$, and 5$d_O$ states into
Ba$_{4f}$ ones. The Ba $L_3$ RIXS spectrum possesses a single peak between 5
and 16 eV, which is formed by the transitions from the {\tg}, O$_{2p}$, and
5$d_O$ states into Ba 5$d$ states. The Ba $K$ RIXS spectrum possesses
relatively small intensity and is formed by the interband transitions from
O$_{2s}$, O$_{2p}$, and 5$d_O$ states into Ba 5$d$, 4$f$, Al 3$p$, and Ir 6$p$
states with almost equal intensity.

The RIXS spectrum of Ba$_5$AlIr$_2$O$_{11}$ at the O $K$ edge consists of
three major inelastic excitations $\le$1 eV, from 1.8 to 4.7 eV, and $>$5
eV. We have found that the first low energy feature is due to the interband
transitions between occupied and empty O$_{\tg}$ states, which appear as a
result of the strong hybridization between oxygen 2$p$ states with Ir {\tg}
LEB and UEB in the vicinity of the Fermi level. The next major structure from
1.8 to 4.7 eV reflects the interband transitions from occupied O 2$p$ states
to empty oxygen states, which originate from the hybridization with Ir {\tg}
states. The O$_{2p}$ $\rightarrow$ {\eg} transitions are situated between 5
and 8 eV, have quite small intensity, and appear in the experimental spectrum
just as a tail. The RIXS as well as XAS spectra at the oxygen $K$ edge in
Ba$_5$AlIr$_2$O$_{11}$ show relatively strong polarization dependence.

We have found much stronger dependence on incident photon energy $E_i$ at the
O $K$ edge than at the Ir $L_3$ edge. With increasing the incident photon
energy, the O $K$ peaks $\le$1 eV and between 1.8 and 4.7 eV are
increased. This occurs in a small energy interval of 1 eV for $E_i$ from 527.6
to 528.6 eV. However, the O$_{2p}$ $\rightarrow$ {\eg} interband transitions,
situated in the 5$-$8 eV energy interval, possess quite small incident photon
energy dependence for both $\sigma$ and $\pi$ polarizations.

\section*{Acknowledgments}

We are thankful to Dr. Alexander Yaresko from the Max Planck Institute FKF in
Stuttgart and Dr. Yuri Kucherenko from the G.V. Kurdyumov Institute for Metal
Physics of the N.A.S. of Ukraine for helpful discussions. 

The studies were supported by the National Academy of Sciences of Ukraine
within the budget program KPKBK 6541230 "Support for the development of
priority areas of scientific research".


\newcommand{\noopsort}[1]{} \newcommand{\printfirst}[2]{#1}
  \newcommand{\singleletter}[1]{#1} \newcommand{\switchargs}[2]{#2#1}

\end{document}